\documentclass[12pt,draftcls, onecolumn]{IEEEtran}
\usepackage{amsmath}
\usepackage{amssymb}
\usepackage{float}
\usepackage{multirow}
\usepackage{graphicx}
\usepackage{color}

\newtheorem{thm}{Theorem}
\newtheorem{cor}{Corollary}
\newtheorem{rem}{Remark}
\newtheorem{defn}{Definition}

\newtheorem{lem}{Lemma}

\floatstyle{ruled}
\newfloat{algorithm}{tbp}{loa}
\providecommand{\algorithmname}{Algorithm}
\floatname{algorithm}{\protect\algorithmname}
\usepackage{algorithm}
\usepackage{algorithmic}

\begin{document}

\title{Convergence Rate of a Message-passing Algorithm for Solving Linear Systems}
\author{{\normalsize{Zhaorong Zhang$^1$, Qianqian Cai$^2$ and Minyue Fu$^{1,2}$, {\em Fellow, IEEE}}} 
\thanks{$^1$School of Electrical Engineering and Computer Science, The University of Newcastle. University Drive, Callaghan, 2308, NSW, Australia.}
\thanks{$^2$School of Automation, Guangdong University of Technology, and Guangdong Key Laboratory of IoT Information Technology, Guangzhou 510006, China.}
\thanks{
This work was supported by the National Natural Science Foundation
of China (Grant Nos. 61633014, 61803101 and U1701264).
E-mails: zhaorong.zhang@uon.edu.au; qianqian.cai@outlook.com; minyue.fu@newcastle.edu.au.}
}
\maketitle
\begin{abstract}
This paper studies the convergence rate of a message-passing distributed algorithm for solving a large-scale linear system. This problem is generalised from the celebrated Gaussian Belief Propagation (BP) problem for statistical learning and distributed signal processing, and this message-passing algorithm is generalised from the well-celebrated Gaussian BP algorithm. Under the assumption of generalised diagonal dominance, we reveal, through painstaking derivations, several bounds on the convergence rate of the message-passing algorithm. In particular, we show clearly how the convergence rate of the algorithm can be explicitly bounded using the diagonal dominance properties of the system. When specialised to the Gaussian BP problem, our work also offers new theoretical insight into the behaviour of the BP algorithm because we use a purely linear algebraic approach for convergence analysis.  
\end{abstract}


\section{Introduction}\label{sec1}

Sparse linear systems are of great interest to many disciplines, and lots of iterative methods exist for solving sparse linear systems; see, e.g., \cite{Horn,Saad,Abur,Khan,Xiao,Weiss,Shental,Mailoutov,Su,Su1,Roy1,Roy2,Wang1,Shi1,Yin}. Distributed solutions are essential for various applications, ranging from sensor networks \cite{Kar,Fu,Fu1}, networked control systems \cite{Morse,Morse1,Wang}, network-based state estimation~\cite{Damian,Russell,Bertrand,Bertrand1}, biological networks \cite{Vicsek1,Vicsek2}, multi-agent systems~\cite{Vicsek3,Lin,Fu2,Fu3}, distributed optimization~\cite{Roy1,Roy2,Nedic,Jak,Lu,1,2,7,8}, consensus and synchronisation \cite{3,6,Fu2020}, and so on.

For a sparse linear system $Ax=b$ with a symmetric and positive definite matrix $A$, a variant  of the well-celebrated Belief Propagation (BP) algorithm \cite{Pearl} called Gaussian Belief Propagation (BP) algorithm \cite{Weiss} can be applied, as such a linear system can be associated with the computation of marginal density functions of a sparse Gaussian graphical model. More specifically, if the joint probability density for a random vector $\mathbf{x}=\mathrm{col}\{\mathbf{x}_1, \mathbf{x}_2, \ldots, \mathbf{x}_n\}$ is given by the following Gaussian density function:
\begin{align}
p(\mathbf{x})\propto \exp\{-\frac{1}{2}\mathbf{x}^TA\mathbf{x}+b^T\mathbf{x}\}, \label{eq:p}
\end{align}
then, the marginal means $x_1, x_2, \ldots, x_n$ for $\mathbf{x}_1, \mathbf{x}_2, \ldots, \mathbf{x}_n$ can be expressed as $Ax=b$. It is well known that the BP algorithm produces correct marginal means in a finite number of iterations when the the corresponding graph (i.e., the Gaussian graphical model) for the joint density function $p(\mathbf{x})$ is acyclic (i.e., no cycles or loops). It is a surprisingly interesting property of the BP algorithm that correct marginal means can be computed asymptotically under appropriate conditions. In \cite{Weiss}, it was shown that Gaussian BP produces asymptotically the correct marginal means under the assumption that the joint information matrix is diagonal dominance. It was relaxed in \cite{Mailoutov} that the same asymptotic convergence holds when the joint information matrix is walk-summable, which is equivalent to the condition of generalised diagonal dominance (see Definition~\ref{def:dd} and Remark~\ref{rem:dd}). In \cite{Su,Su1},  necessary and sufficient conditions for asymptotic convergence of the Gaussian BP algorithm are studied. An upper bound on its convergence rate is given in \cite{Roy1} under a somewhat different diagonal dominance condition. These results promise excellent use of the Gaussian BP algorithm when the matrix $A$ is symmetric and positive definite. Also see \cite{14} for applications in distributed optimisation. 

For a general sparse linear system with a non-symmetric matrix $A$, solving $Ax=b$ via Gaussian BP can be done in two ways: either solving $(A^TA)x=A^Tb$ or solving $(AA^T)y=b$ and computing $x=A^Ty$. It is easy to see that for a full rank matrix $A$, either $A^TA$ or $AA^T$ is symmetric and positive-definite. But this approach would involve substantially more computational complexity in comparison to solving $Ax=b$ with a symmetric and positive definite $A$. This is due to the fact that if the graph for $A$ has $m$ edges, the graph for $A^TA$ or $AA^T$ has roughly $m^2$ edges. Our earlier paper \cite{Fu5} generalises the Gaussian BP algorithm to a similar message-passing distributed algorithm (see details of the algorithm in Section~\ref{sec3}). It has been shown in \cite{Fu5} that this algorithm enjoys similar properties of the Gaussian BP algorithm. Namely, if an induced graph (similar to the Gaussian graphical model for Gaussian distributions) is acyclic, the algorithm converges in a finite number of iterations with the correct distributed solution for $x$. For a general (cyclic) induced graph, under the assumption that the matrix $A$ satisfies a walk-summability condition, the algorithm also converges asymptotically to the correct $x$.  

The purpose of this paper is to study the convergence rate of the message-passing algorithm in \cite{Fu5}.  We note that various convergence results can be found in \cite{Weiss,Mailoutov,Roy1,Roy2} for the Gaussian BP algorithm, but the analyses in these references all rely on the Gaussian graphical model for the underlying optimisation problem. Unfortunately, this property breaks down when the matrix $A$ is not symmetric and positive definite. In order to carry out convergence analysis for the general case, we generalise the analysis tools in two key references (\cite{Weiss} and \cite{Mailoutov}) for the Gaussian BP algorithm. Instead of relying on the Gaussian graphical model, we use a basic linear algebraic approach to characterise the convergence rate of the message-passing algorithm through painstaking derivations.    The contributions of this paper are summarised below.
\begin{itemize}
\item Our first main result (Theorem~\ref{thm:1}) gives an explicit bound for the convergence rate of the message-passing algorithm in \cite{Fu5}.  This bound relates the convergence rate to the diagonal dominance parameters and the topology of the network graph, clearly revealing how the messages pass through the graph as the iterative solutions evolve.  The knowledge of such explicit bound is very important in determining the number of iterations required to reach a given level of accuracy, and in understanding the computational complexity of the algorithm. 
\item A direct implication of the main result above is a simple bound for the convergence rate using the spectral radius of a matrix related to matrix $A$ (Corollary \ref{cor:1}). This bound is known for the symmetric case of $A$, but we have shown that the same bound holds in the general case. 
\item We also analyse the asymptotic convergence behaviour of the message-passing algorithm and reveals its close relationship with the so-called loop gain of each loop in the graph. Through this relationship, another bound is given for the asymptotic convergence rate (Corollary \ref{cor:2}).  
\item Our results generalise the convergence rate results on the Gaussian BP algorithm for the case of symmetric $A$, including the important results in \cite{Weiss,Mailoutov,Roy1}. More importantly, our work also offers new theoretical insights into the behaviour of the BP algorithm because we use a purely linear algebraic approach for convergence analysis, whereas previous analysis results are all based on the Gaussian random field interpretation of the algorithm, i.e., they focus on tracking the Gaussian means and variances of the marginal distributions which have no counterparts for a general linear system. 
\end{itemize}

The rest of the paper is organised as follows. Section~\ref{sec2} formulates the distributed linear system problem; Section~\ref{sec3} introduces the message-passing distributed algorithm; Section~\ref{sec4} carries out preliminary analysis for its convergence; Section~\ref{sec5} characterises the convergence rate; Section~\ref{sec6} provides several illustrating examples; and Section~\ref{sec7} concludes the paper. 

\section{Problem Formulation}\label{sec2}

\subsection{Problem Formulation}
Consider a network of nodes $(1,2,\ldots, n)$ associated with a state vector $x=\mathrm{col}\{x_1,x_2,\ldots, x_n\}\in \mathbb{R}^n$, where $x_i\in \mathbb{R}$ is the unknown variable for node $i$. The information available at each node $i$ is that $x$ satisfies a linear system:
$$a_i^T x = b_i,$$ 
where $a_i=\mathrm{col}\{a_{i1}, a_{i2}, \ldots\, a_{in}\}\in \mathbf{R}^{n}$ is a column vector and $b_i$ is a scalar, i.e., the values of $a_i$ and $b_i$ are local information known only to node $i$.  Collectively, the common state satisfies 
\begin{align}
A x &= b \label{eq:Axb}
\end{align}
with $A=\mathrm{col}\{a_1^T, a_2^T, \ldots, a_n^T\}$ and $b=\mathrm{col}\{b_1, b_2, \ldots, b_n\}$.  

This paper will focus on a class of linear systems (\ref{eq:Axb}) for which the matrix $A$ satisfies the so-called {\em generalised diagonal dominance condition, which is to be defined later. A special property of this condition is that $A$ is invertible.}  Denote the solution of (\ref{eq:Axb}) by 
\begin{align}
x^{\star}&=A^{-1}b. \label{xstar}
\end{align} 

Define the {\em induced graph} $G=\{V, E\}$ with $V=\{1, 2, \ldots, n\}$ and $E=\{(i,j)|a_{ij}\ne 0 \mathrm{\ or\ } a_{ji}\ne0\}$. We associate each node $i$ with $(a_{ii}, b_i)$,  and each edge $(i, j)$ with $a_{ij}$. Note in particular that $G$ is {\em undirected}, i.e., $(i,j)\in E$ if and only if $(j,i)\in E$.  For each node $i\in V$, we define its {\em neighbouring set} as $N_i=\{j|(i,j)\in E\}$ and denote its {\em cardinality} by $|N_i|$. We assume in the sequel that $|N_i|\ll n$ for all $i\in V$. The graph $G$ is said to be {\em connected} if for any $i, j\in V$, there exists a (connecting) {\em path} of  $(i, i_1), (i_1, i_2), \ldots, (i_k, j)\in E$. Such a path is denoted by $\{i, i_1, i_2, \ldots, i_k,j\}$.  The {\em length} of a path equals the number of connecting edges, with the convention that the length of a single node is zero. The {\em distance} between two nodes is the minimum length of a path connecting the two nodes. It is obvious that a graph with a finite number of nodes is either connected or composed of a finite number of disjoint subgraphs with each of them being a connected graph. The {\em diameter} of a connected graph is defined to be largest distance between two nodes in the graph. The diameter of a disconnected graph is the largest diameter of a connected subgraph. By this definition, the diameter of a graph with a finite number of nodes (or a finite graph) is always finite.  A {\em loop} is defined to be a path starting and ending at node $i$ through a node $j\ne i$. A graph is said to be {\em acyclic} if it does not contain any loop. A {\em cyclic} (or {\em loopy}) graph is a graph with at least one loop.  

The {\em distributed linear system problem} we are interested in is to devise an iterative algorithm for each node $i\in V$ to execute so that node $i$ will be able to solve $x_i$.  Note that in this problem formulation, node $i$ is only interested in its local variable $x_i$, and not interested in knowing the solution of $x_j$ for any other node $j\in V$. This is {\em sharply different} from many distributed methods in the literature which require each node to compute the whole solution of $x$. For a large network, computing the whole solution of $x$ is not only burdensome for each node, but also unnecessary in most applications. 

We want the algorithm to be of low complexity and fast convergence.   Certain {\em constraints} need to be imposed on the algorithm's complexities of communication, computation and storage to call it {\em distributed}. In our paper, these include:
\begin{enumerate}
\item[C1:] Local information exchange:  Each node $i$ can exchange information with each $j\in N_i$ only once per iteration. 
\item[C2:] Local computation: Each node $i$'s computational load should be at most $O(|N_i|)$ per iteration. 
\item[C3:] Local storage: Each node $i$'s storage should be at most $O(|N_i|)$ over all iterations. 
\end{enumerate}


\begin{defn}\label{def:dd}
\cite{Horn,Saad} A matrix $A=\{a_{ij}\}\in \mathbb{R}^{n\times n}$ is said to be {\em strictly diagonally dominant} (or simply {\em diagonally dominant} in this paper) if $a_{ii}>0$ and $a_{ii}>\sum_{j\ne i}|a_{ij}|$ for every $i\in V$.  The matrix $A$ is said to be {\em generalised diagonally dominant} if there exists a diagonal matrix $D=\mathrm{diag}\{d_i\}$ with all $d_i>0$ such that $D^{-1}AD$ is diagonally dominant. 
\end{defn}
  
\begin{rem}\label{rem:dd}
It is known \cite{Mailoutov} that $A$ with $a_{ii}=1$ for all $i$ is generalised diagonally dominant if and only if $R=I-A$ is so-called {\em walk summable}, a notion very important in the convergence analysis of Gaussian belief propagation algorithm (see \cite{Mailoutov} for more detailed connection between diagonal dominance and walk summability).  
\end{rem}

\section{Distributed Solver for Linear systems}\label{sec3}

The distributed algorithm for solving (\ref{eq:Axb}) to be studied in this paper is listed in Algorithm~\ref{alg1}. This was proposed in \cite{Fu5} and was generalised from the well-known Gaussian belief propagation algorithm \cite{Weiss} corresponding to a symmetric and positive definite matrix $A$.  

In each iteration $k$ of Algorithm~\ref{alg1}, each node $i$ computes variables $a_{i\rightarrow j}^{(k)}$ and $b_{i\rightarrow j}^{(k)}$ for each of its neighboring node $j\in N_i$ and transmit them to node $j$.  All the nodes execute the same algorithm concurrently.  

\begin{algorithm}[ht] 
\protect\protect\protect\protect\protect\protect\protect\caption{(Distributed Solver for Linear systems)}
\label{alg1} \begin{itemize}
\item \textbf{Initialization:} For each node $i$, do: For each $j\in N_i$, set $a_{i\rightarrow j}^{(0)} = a_{ii}, b_{i\rightarrow j}^{(0)}=b_i$ and transmit them to node $j$.
\item \textbf{Main loop:} At iteration $k=1,2,\cdots$, for each node $i$, compute
\begin{align}
a_i^{(k)} &= a_{ii} - \sum_{v\in N_i} \frac{a_{vi}a_{iv}}{a_{v\rightarrow i}^{(k-1)}} \label{ak0}\\
b_i^{(k)} &= b_i-\sum_{v\in N_i} \frac{a_{iv}b_{v\rightarrow i}^{(k-1)}}{a_{v\rightarrow i}^{(k-1)}}
\label{bk0}\\
x_i^{(k)}& = \frac{b_i^{(k)}} {a_i^{(k)}}, \label{xik}
\end{align}
then for each $j\in N_i$, compute
\begin{align}
a_{i\rightarrow j}^{(k)} &= a_i^{(k)}+ \frac{a_{ji}a_{ij}}{a_{j\rightarrow i}^{(k-1)}} \label{ak}\\
b_{i\rightarrow j}^{(k)} &= b_i^{(k)} + \frac{a_{ij}b_{j\rightarrow i}^{(k-1)}}{a_{j\rightarrow i}^{(k-1)}} \label{bk}
\end{align}
and transmit them to node $j$. 
\end{itemize}
\end{algorithm}

\section{Convergence Analysis for Loopy Graphs}\label{sec4}

In this section, we introduce a slightly different diagonal dominance notion to study the convergence properties of the BP algorithm for loopy graphs.  It suffices to consider a connected graph $G$, which we will assume in the rest of the paper. 

\begin{defn}\label{def:dd1}
For a given $D=\mathrm{diag}\{d_i\}$ with all $d_i>0$, the matrix $A$ is said to be {\em $D$-scaled diagonally dominant} \cite{Roy1} if 
$a_{ii}>0$ and $\varrho_i<1$ for every $i\in V$, where
\begin{align}
\varrho_i=\frac{\sum_{j\ne i}|a_{ij}|d_j}{a_{ii}d_i}. \label{eq:rhoi}
\end{align}
The matrix $A$ is said to be {\em weakly $D$-scaled diagonally dominant} if $a_{ii}>0$ for every $i\in V$ and $\varrho_i\varrho_j< 1$ for every $(i,j)\in E$ (Note that it is not necessary to have all $\varrho_i<1$).
\end{defn}

\begin{rem}\label{rem:dd0}
It is clear from the definitions above that $D$-scaled diagonal dominance implies generalised diagonal dominance because $D^{-1}AD$ is diagonally dominant, but a generalised diagonally dominant matrix $A$ may require a different diagonal matrix $\Delta$ such that $\Delta^{-1}A\Delta$ is diagonally dominant. Searching for such a $\Delta$ amounts to a separate optimisation problem. In this sense, assuming $D$-scaled diagonal dominance for a given $D$ is somewhat stronger than assuming generalised diagonal dominance. Similarly, assuming weakly $D$-scaled diagonal dominance for a given $D$ is weaker than assuming $D$-scaled diagonal dominance. But as we will show in Appendix (Lemma~\ref{lem:d0}) that weakly $D$-scaled diagonal dominance also implies generalised diagonal dominance (but for a possibly different diagonalising matrix). Denoting by $\mathcal{A}_{\mathrm{dd}}, \mathcal{A}_{\mathrm{ddd}}, \mathcal{A}_{\mathrm{wddd}}$ and $\mathcal{A}_{\mathrm{gdd}}$ the sets of $n\times n$ matrices which are diagonally dominant, $D$-scaled diagonally dominant, weakly $D$-scaled diagonally dominant and generalised diagonally dominant, respectively, the above implies
$\mathcal{A}_{\mathrm{dd}}\subset \mathcal{A}_{\mathrm{ddd}}\subset \mathcal{A}_{\mathrm{wddd}}\subset \mathcal{A}_{\mathrm{gdd}}$ and $\cup_{D} \mathcal{A}_{\mathrm{ddd}} = \mathcal{A}_{\mathrm{gdd}}$. 
\end{rem}

\subsection{A Useful Property}

\begin{lem}\label{lem:3}
Suppose $A \in \mathbb{R}^{n\times n}$ is weakly $D$-scaled diagonally dominant for a given positive diagonal matrix $D$. Then, for all node $i\in V$, $j\in N_i$ and $k\ge0$, it holds that 
 \begin{align}
\varrho_ia_{i\rightarrow j}^{(k)}d_i  &\ge |a_{ij}|d_j. \label{eq:aij_bound}
\end{align}
\end{lem}

\begin{IEEEproof}
We proceed by induction on the iteration number $k$. It is clear from (\ref{eq:rhoi}) that, for any  node $i\in V$ and $j\in N_i$, we have $\varrho_ia_{i\rightarrow j}^{(0)}d_i=\varrho_ia_{ii}d_i \ge  |a_{ij}|d_j$, which confirms (\ref{eq:aij_bound}) for $k=0$.  
We now assume that,  for all $i\in V$ and $j\in N_i$, we have $\varrho_ia_{i\rightarrow j}^{(k-1)}d_i\ge |a_{ij}|d_j$ for some $k\ge 1$. We claim that (\ref{eq:aij_bound}) holds for $k$.  Indeed, from Algorithm~\ref{alg1}, we have 
\begin{align*}
a_{i\rightarrow j}^{(k)} &= a_{ii} - \sum_{v\in N\backslash j} \frac{a_{iv}a_{vi}}{a_{v\rightarrow i}^{(k-1)}}.
\end{align*}
By Definition~\ref{def:dd1} and the induction assumption above, we get
\begin{align*}
\varrho_ia_{i\rightarrow j}^{(k)}d_i&
\ge |a_{ij}|d_j+\sum_{v\in N_i\backslash j} \left (|a_{iv}|d_v - \varrho_i\frac{|a_{iv}a_{vi}|}{a_{v\rightarrow i}^{(k-1)}}d_i\right )\\
&\ge |a_{ij}|d_j + \sum_{v\in N_i \backslash j} \left (
|a_{iv}|d_v - \varrho_i\frac{\varrho_vd_v|a_{iv}a_{vi}|}{|a_{vi}|}\right )\\
&\ge |a_{ij}|d_j + \sum_{v\in N_i \backslash j} 
|a_{iv}|d_v(1 -\varrho_i\varrho_{v})\\
&\ge |a_{ij}|d_j. 
\end{align*}
So our claim holds. By induction, (\ref{eq:aij_bound}) holds for all $k\ge0$.  
\end{IEEEproof}

\subsection{Unwrapped Tree}

The convergence analysis of Algorithm~\ref{alg1} relies critically on the concept of unwrapped tree (also known as {computation tree} \cite{Roy1,Roy2}). This reliance was firstly demonstrated in \cite{Weiss} for the Gaussian BP case (i.e., with a symmetric $A$), and we generalise this approach to the case with a general matrix $A$. 

Following the work of~\cite{Weiss}, we construct an {\em unwrapped tree} with {\em depth} $k>0$ for a loopy graph $G$~\cite{Weiss}. Take any node $r\in V$ to be the root and then iterate the following procedure $k$ times:
\begin{itemize}
\item Find all leaves of the tree (start with the root);
\item For each leaf, find all the nodes in the loopy graph that neighbour this leaf node, except its parent node in the tree, and add all these nodes as the children to this leaf node.
\end{itemize}

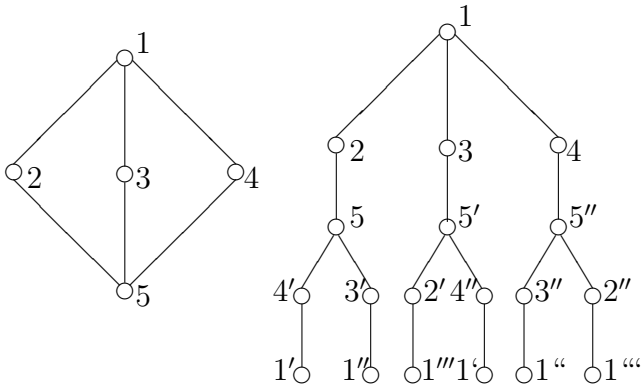
\begin{figure}[ht]
\begin{picture}(240,135)
\put(54,124){\circle{6}}
\put(57,123){\line(1,-1){39}} 
\put(51,123){\line(-1,-1){39}} 
\put(54,121){\line(0,-1){38}} 
\put(54,80){\circle{6}}
\put(12,81){\circle{6}}
\put(96,81){\circle{6}}
\put(54,77){\line(0,-1){38}}
\put(56,38){\line(1,1){40}} 
\put(52,38){\line(-1,1){40}} 
\put(54,36){\circle{6}}
\put(58,30){$5$}
\put(58,126){$1$}
\put(58,75){$3$}
\put(17,75){$2$}
\put(99,75){$4$}

\put(176,134){\circle{6}}
\put(179,133){\line(1,-1){39}} 
\put(173,133){\line(-1,-1){39}} 
\put(176,131){\line(0,-1){38}} 
\put(176,90){\circle{6}}
\put(134,91){\circle{6}}
\put(218,91){\circle{6}}
\put(176,87){\line(0,-1){25}}
\put(134,88){\line(0,-1){25}}
\put(218,88){\line(0,-1){25}}
\put(134,60){\circle{6}}
\put(176,60){\circle{6}}
\put(218,60){\circle{6}}
\put(180,136){$1$}
\put(180,85){$3$}
\put(139,85){$2$}
\put(221,85){$4$}
\put(139,60){$5$}
\put(180,60){$5'$}
\put(222,60){$5''$}
\put(133,57){\line(-3,-5){12}}
\put(135,57){\line(3,-5){12}}
\put(175,57){\line(-3,-5){12}}
\put(177,57){\line(3,-5){12}}
\put(217,57){\line(-3,-5){12}}
\put(219,57){\line(3,-5){12}}
\put(121,34){\circle{6}}
\put(147,34){\circle{6}}
\put(163,34){\circle{6}}
\put(190,34){\circle{6}}
\put(205,34){\circle{6}}
\put(231,34){\circle{6}}
\put(121,31){\line(0,-1){24}}
\put(147,31){\line(0,-1){24}}
\put(163,31){\line(0,-1){24}}
\put(190,31){\line(0,-1){24}}
\put(205,31){\line(0,-1){24}}
\put(231,31){\line(0,-1){24}}
\put(121,4){\circle{6}}
\put(147,4){\circle{6}}
\put(163,4){\circle{6}}
\put(190,4){\circle{6}}
\put(205,4){\circle{6}}
\put(231,4){\circle{6}}
\put(110,32){$4'$}
\put(137,32){$3'$}
\put(167,32){$2'$}
\put(177,32){$4''$}
\put(209,32){$3''$}
\put(235,32){$2''$}
\put(110,2){$1'$}
\put(136,2){$1''$}
\put(166,2){$1'''$}
\put(179,2){$1`$}
\put(209,2){$1``$}
\put(235,2){$1```$}
\end{picture}
  \caption{Left: A loopy graph. Right: The unwrapped tree for root node 1 with 4 layers ($k=4$)}\label{fig:2}
\end{figure}

The variables and weights for each node in the unwrapped tree are copied from the corresponding nodes in the loopy graph. It is clear that taking each node as root node will generate a different unwrapped tree.  Fig.~\ref{fig:2} shows the unwrapped tree around root node 1 for a loopy graph. Note, for example, that nodes $1', 1'',1''', 1`, 1``,1```$ all carry the same values $b_1$ and $a_{11}$. Similarly, if node 1' is the parent (or child) of node $j'$ in the unwrapped tree, then $a_{1'j'}=a_{1j}$ and $a_{j'1'}=a_{j1}$. 

Without loss of generality, we will take the root node $r=1$ in the sequel.  Denote the unwrapped tree as $\breve{G}=\{\breve{V}, \breve{E}\}$ with the associated matrix $\breve{A}$ and vector $\breve{b}$.  Also denote by the node mapping from $\breve{G}$ to $G$ as $\sigma(\cdot)$, i.e., a node $i$ in $\breve{G}$ is mapped to node $\sigma(i)$ in $G$. But whenever there is no confusion, we do not differentiate $i$ and $\sigma(i)$ for notational convenience.  It is obvious that $\breve{G}$ is connected by construction.  The linear system corresponding to the unwrapped graph is described by
\begin{align}
\breve{A}\breve{x} &= \breve{b} \label{eq:unwrapped}
\end{align}
with the following: If $i$ is an interior (non-leaf) node of $\breve{G}$, then the $i$-th row of (\ref{eq:unwrapped}) is
\begin{align*}
a_{ii}\breve{x}_i+\sum_{u\in \breve{N}_i} a_{iu}\breve{x}_u &= b_i, 
\end{align*}
and if $i$ is a leaf node of $\breve{G}$ with parent node $j$, then the $i$-th row of (\ref{eq:unwrapped}) is 
\begin{align*} 
a_{ii}\breve{x}_i+a_{ij}\breve{x}_j &=b_i. 
\end{align*}
Denote by $\breve{x}^{\star}$ the solution to (\ref{eq:unwrapped}). We have the following properties, generalised from \cite{Weiss} for diagonal dominance matrices. 

\begin{lem}\label{lem:unwrapped}
Suppose $A$ is weakly $D$-scaled diagonally dominant for some positive diagonal matrix $D=\mathrm{diag}\{d_i\}$. Then, $\breve{A}$ is weakly $\breve{D}$-scaled diagonally dominant with unwrapped $\breve{D}=\mathrm{diag}\{\breve{d}_i\}$ defined by $\breve{d}_i=d_{\sigma(i)}$. Moreover, applying Algorithm~\ref{alg1} for $k$ iterations to $G$ or $\breve{G}$ yields the same results for the root node, i.e., $x_1^{(k)}=\breve{x}_1^{(k)}$.
\end{lem}

\begin{IEEEproof}
For any interior node $i$, it is obvious that the $i$-th row of $\breve{A}$ has the same diagonal dominance property as the $\sigma(i)$-th row of (\ref{eq:Axb}). Now consider any leaf node $i$ with parent node $j$. It is also obvious that the diagonal dominance property of the $i$-th row of (\ref{eq:unwrapped}) is implied by that of the $\sigma(i)$-th row of (\ref{eq:Axb}) because the former has only one off-diagonal term left. The property that $x_1^{(k)}=\breve{x}_1^{(k)}$ follows from the construction of the unwrapped tree \cite{Weiss}. 
\end{IEEEproof}

\section{Convergence Rate of Algorithm \ref{alg1}}\label{sec5}

In this section, we provide our main result (Theorem~\ref{thm:1}) to give a general characterisation for the convergence rate of Algorithm~\ref{alg1} under the weakly $D$-scaled diagonal dominance assumption.  This characterisation shows how the convergence of the estimates $x_i^{(k)}$ in each iteration are related to the monotonic decreasing of certain internal variables ($\eta_{i\rightarrow j}^{(k)}$ to be defined below), but the expression is technical and difficult to be applied directly. For this reason, this result is then specialised to give more insightful conditions for the convergence rate (Theorem~\ref{thm:2} and Corollaries~\ref{cor:1}-\ref{cor:2}).
 
\subsection{General Characterisation of Convergence Rate}

Our main result below is established based on the convergence analysis in the previous section. 

\begin{thm}\label{thm:1}
Suppose $A\in \mathbb{R}^{n\times n}$ is weakly $D$-scaled diagonally dominant for some diagonal matrix $D>0$. Then, 
\begin{align}
\frac{|x_i^{(k)} - x_i^{\star}|}{d_i} &\le \varrho_i \frac{\sum_{v\in N_i} \Lambda_{v\rightarrow i}^{(k-1)}\eta_{v\rightarrow i}^{(k-1)}}{\sum_{v\in N_i}\Lambda_{v\rightarrow i}^{(k-1)}}\|x^{\star}\|_d
\label{eq:rate}
\end{align}
for all $i\in V$ and $k>0$, where $\|x\|_d=\max_v|x_v|d_v^{-1}$ is the {\em scaled max norm}, and $\Lambda_{i\rightarrow j}^{(\ell)}$ and $\eta_{i\rightarrow j}^{(\ell)}$ are defined as follows:  For every $i\in V$ and $j\in N_i$, 
\begin{align}
\Lambda_{i\rightarrow j}^{(\ell)} &= |a_{ji}|d_i - \frac{a_{ji}a_{ij}d_j\varrho_j}{a_{i\rightarrow j}^{(\ell)}} > 0,   \ \ \forall \ell\ge0, \label{eq:beta1}\\
\eta_{i\rightarrow j}^{(0)}&=\varrho_i, \nonumber \\
\eta_{i\rightarrow j}^{(\ell)}&=\varrho_i \frac{\sum_{v\in N_i\backslash j} \Lambda_{v\rightarrow i}^{(\ell-1)} \eta_{v\rightarrow i}^{(\ell-1)}}{|a_{ij}|d_j(1-\varrho_i\varrho_j)+\sum_{v\in N_i\backslash j} \Lambda_{v\rightarrow i}^{(\ell-1)}},   \forall \ell>0.
 \label{eq:beta_l}
\end{align}
\end{thm}

\begin{IEEEproof}
Without loss of generality, we prove  (\ref{eq:rate}) for node $i=1$ only.  Construct the unwrapped tree $\breve{G}$ with depth $k$ as discussed before with node 1 as the root node. Consider the following linear system
\begin{align}
\breve{A}\vec{x} &= \vec{b},\label{eq:unwrapped1}
\end{align}
where $\vec{b}$ is modified from $\breve{b}$ such that, for any leaf node $i$ with parent node $j$,
\begin{align*}
\vec{b}_i &= b_i-\sum_{v\in N_i\backslash j} a_{iv} x_v^{\star}.
\end{align*}
By construction, it is clear that $x^{\star}$ satisfies (\ref{eq:unwrapped1}). Since $\breve{A}$ is generalised diagonally dominant, it is invertible and thus $x^{\star}$ is the unique solution to (\ref{eq:unwrapped1}). 

Now consider the next linear system:
\begin{align}
\tilde{A}\mathbf{x} &= \mathbf{b}, \label{eq:unwrapped2}
\end{align}
where $\mathbf{b}=\breve{b}-\vec{b}$. By construction, we have $\mathbf{b}_i=0$ for every interior node $i$ of $\breve{G}$ and 
\begin{align}
\mathbf{b}_i&=\sum_{v\in N_i\backslash j} a_{iv}x_v^{\star} \label{eq:bi2}
\end{align}
for every leaf node $i$ with parent node $j$.

Applying Algorithm~\ref{alg1} to (\ref{eq:unwrapped2}) for $k$ iterations, we have, from Lemma~\ref{lem:unwrapped}, that
\begin{align*}
x_1^{(k)}-x_1^{\star} &= \breve{x}_1^{(k)}-\vec{x}_1^{(k)} = \mathbf{x}^{(k)}.
\end{align*}
Hence, it suffices to bound $\mathbf{x}^{(k)}$. This is done by tracking $a_{i\rightarrow j}^{(\ell)}$ and $\mathbf{b}_{i\rightarrow j}^{(\ell)}$ for $\ell=1,2,\ldots, k$. 

Instead of tracking $\mathbf{b}_{i\rightarrow j}^{(\ell)}$ directly, we consider its scaled version below:
\begin{align}
\beta_{i\rightarrow j}^{(\ell)} &= \frac{a_{ji}}{a_{i\rightarrow j}^{(\ell)}\Lambda_{i\rightarrow j}^{(\ell)}}\mathbf{b}_{i\rightarrow j}^{(\ell)} \label{eq:beta}
\end{align}
with $\Lambda_{i\rightarrow j}^{(\ell)}$ defined in (\ref{eq:beta1}).  To show 
$\Lambda_{i\rightarrow j}^{(\ell)}>0$, we note that $a_{ji}\ne0$ (because $j$ is the parent node of $i$) and it follows from Lemma~\ref{lem:3} that 
\begin{align*}
\Lambda_{i\rightarrow j}^{(\ell)} &= |a_{ji}|d_i - \frac{a_{ji}a_{ij}d_j\varrho_j}{a_{i\rightarrow j}^{(\ell)}} \\
&\ge |a_{ji}|d_i - \frac{|a_{ji}a_{ij}|\varrho_j\varrho_id_i}{|a_{ij}|}\\
&=|a_{ji}|d_i (1-\varrho_j\varrho_i)>0.
\end{align*} 
Moreover, we have the following key property:
\begin{align}
\frac{a_{i\rightarrow j}^{(\ell)}\Lambda_{i\rightarrow j}^{(\ell)}}{|a_{ji}|}&= a_{i\rightarrow j}^{(\ell)}d_i -\frac{a_{ij}a_{ji}d_j\varrho_j}{|a_{ji}|}\nonumber \\
&=a_{ii}d_i-\sum_{v\in N_i\backslash j}\frac{a_{iv}a_{vi}d_i}{a_{v\rightarrow i}^{(\ell-1)}} - \frac{a_{ij}a_{ji}d_j\varrho_j}{|a_{ji}|}\nonumber  \\
&=\left (\varrho_i^{-1}|a_{ij}|d_j-\frac{a_{ij}a_{ji}d_j\varrho_j}{|a_{ji}|}\right ) \nonumber \\
&\ \ + \sum_{v\in N_i\backslash j}\left (\varrho_i^{-1}|a_{iv}|d_v-\frac{a_{iv}a_{vi}d_i}{a_{v\rightarrow i}^{(\ell-1)}}\right ) \nonumber \\
&\ge \varrho_i^{-1}|a_{ij}|d_j(1-\varrho_i\varrho_j)+
\varrho_i^{-1}  \sum_{v\in N_i\backslash j} \Lambda_{v\rightarrow i}^{(\ell-1)}. \label{eq:Lambdaij}
\end{align}

Start from the $k$-th layer. For any node $i$ in the $k$-th layer, denote by $j$ its parent node. To bound $\beta_{i\rightarrow j}^{(0)}$, we have
\begin{align*}
|\beta_{i\rightarrow j}^{(0)}| &= \frac{|a_{ji}|}{a_{i\rightarrow j}^{(0)}\Lambda_{i\rightarrow j}^{(0)}}|\mathbf{b}_{i\rightarrow j}^{(0)}|\\
&=\frac{|a_{ji}|}{a_{ii}|a_{ji}|d_i-a_{ji}a_{ij}d_j\varrho_j}|\sum_{v\in N_i\backslash j}a_{iv}x_v^{\star}|\\
&\le \frac{1}{a_{ii}d_i-|a_{ij}|d_j\varrho_j}\sum_{v\in N_i\backslash j}|a_{iv}|d_v\|x^{\star}\|_d\\
&\le \frac{\sum_{v\in N_i\backslash j}|a_{iv}|d_v}{\varrho_i^{-1}|a_{ij}|d_j +\varrho_i^{-1}\sum_{v\in N_i\backslash j} |a_{iv}|d_v-|a_{ij}|d_j\varrho_j}\|x^{\star}\|_d\\
&\le \frac{\sum_{v\in N_i\backslash j}|a_{iv}|d_v}{\varrho_i^{-1}\sum_{v\in N_i\backslash j} |a_{iv}|d_v}\|x^{\star}\|_d\\
&=\varrho_i\|x^{\star}\|_d=\eta_{i\rightarrow j}^{(0)}\|x^{\star}\|_d.
\end{align*}

Now we move to the $(k-1)$-th layer. For any node $i$ in the $(k-1)$-th layer, denote by $j$ its parent node, and $v\in N_i\backslash j$ its children.  Note that $\mathbf{b}_i=0$, thus,
\begin{align*}
|\beta_{i\rightarrow j}^{(1)}|&=\left |\frac{a_{ji}}{a_{i\rightarrow j}^{(1)}\Lambda_{i\rightarrow j}^{(1)}}\mathbf{b}_{i\rightarrow j}^{(1)}\right | \\
&=\left |-\frac{a_{ji}}{a_{i\rightarrow j}^{(1)}\Lambda_{i\rightarrow j}^{(1)}}\sum_{v\in N_i\backslash j}\frac{a_{iv}}{a_{v\rightarrow i}^{(0)}}\mathbf{b}_{v\rightarrow i}^{(0)}\right |\\
&=\left |\frac{a_{ji}}{a_{i\rightarrow j}^{(1)}\Lambda_{i\rightarrow j}^{(1)}}\sum_{v\in N_i\backslash j}\Lambda_{v\rightarrow i}^{(0)}\beta_{v\rightarrow i}^{(0)}\right |\\
&\le \frac{|a_{ji}|}{a_{i\rightarrow j}^{(1)}\Lambda_{i\rightarrow j}^{(1)}}\sum_{v\in N_i\backslash j}\Lambda_{v\rightarrow i}^{(0)}\eta_{v\rightarrow i}^{(0)} \|x^{\star}\|_d.
\end{align*}
The last step above used the bound on $\beta_{v\rightarrow i}^{(0)}$. Using (\ref{eq:Lambdaij}), we further obtain 
\begin{align*}
|\beta_{i\rightarrow j}^{(1)}|&\le \varrho_i\frac{\sum_{v\in N_i\backslash j}\Lambda_{v\rightarrow i}^{(0)}\eta_{v\rightarrow i}^{(0)}}{|a_{vi}|d_i(1-\varrho_v\varrho_i)+\sum_{v\in N_i\backslash j} \Lambda_{v\rightarrow i}^{(0)} } \|x^{\star}\|_d\\
&=\eta_{i\rightarrow j}^{(1)}\|x^{\star}\|_d.
\end{align*}

The process above can be repeated until we get to the first layer for which every node $i$ has node 1 as its parent, and that 
\begin{align*}
|\beta_{i\rightarrow 1}^{(k-1)}| &\le \varrho_i\frac{\sum_{v\in N_i\backslash 1}\Lambda_{v\rightarrow i}^{(k-2)}\eta_{v\rightarrow i}^{(k-2)}}{|a_{vi}|d_i(1-\varrho_v\varrho_i)+\sum_{v\in N_i\backslash 1}\Lambda_{v\rightarrow i}^{(k-2)}} \|x^{\star}\|_d\\
&= \eta_{i\rightarrow 1}^{(k-1)}\|x^{\star}\|_d.
\end{align*}
Finally,  apply (\ref{bk0})-(\ref{xik}) to compute the following for node 1:
\begin{align*}
\mathbf{b}_1^{(k)} &=-\sum_{v\in N_1} \frac{a_{1v}\mathbf{b}_{v\rightarrow 1}^{(k-1)}}{a_{v\rightarrow 1}^{(k-1)}} =- \sum_{v\in N_1}\Lambda_{v\rightarrow 1}^{(k-1)}\beta_{v\rightarrow 1}^{(k-1)}, \\
\mathbf{x}_1^{(k)} &= \frac{\mathbf{b}_1^{(k)}}{a_{1}^{(k)}} =-\frac{1}{a_{1}^{(k)}}\sum_{v\in N_1} \Lambda_{v\rightarrow 1}^{(k-1)} \beta_{v\rightarrow 1}^{(k-1)}.
\end{align*}
Using (\ref{ak0}), we have 
\begin{align*}
a_{1}^{(k)} &= a_{11} - \sum_{v\in N_1} \frac{a_{1v}a_{v1}}{a_{v\rightarrow 1}^{(k-1)}}\\ &=d_1^{-1}\varrho_1^{-1}\sum_{v\in N_1}|a_{1v}|d_v- \sum_{v\in N_1} \frac{a_{1v}a_{v1}}{a_{v\rightarrow 1}^{(k-1)}}\\
&=d_1^{-1}\varrho_1^{-1}\sum_{v\in N_1}\Lambda_{v\rightarrow 1}^{(k-1)}.
\end{align*}
This leads to 
\begin{align*}
\frac{|\mathbf{x}_1^{(k)}|}{d_1} &\le \varrho_1\frac{\sum_{v\in N_1} \Lambda_{v\rightarrow 1}^{(k-1)} |\beta_{v\rightarrow 1}^{(k-1)}|}{\sum_{v\in N_1}\Lambda_{v\rightarrow 1}^{(k-1)}} \\
 &\le \varrho_1\frac{\sum_{v\in N_1} \Lambda_{v\rightarrow 1}^{(k-1)} \eta_{v\rightarrow 1}^{(k-1)}}{\sum_{v\in N_1}\Lambda_{v\rightarrow 1}^{(k-1)}} \|x^{\star}\|_d.
\end{align*}

Noting $\mathbf{x}_1^{(k)}=x_1^{(k)}-x_1^{\star}$ and that the root node is arbitrary, we conclude that (\ref{eq:rate}) holds for all $i\in V$ and $k\ge0$. 
\end{IEEEproof}

\subsection{More Direct Convergence Rate Characterisations}

Next, we apply Theorem~\ref{thm:1} to provide more direct characterisations of the convergence rate for Algorithm~\ref{alg1}. 

We first show that the convergence properties of Algorithm~\ref{alg1} are invariant under the diagonal transformation of $A$. 
\begin{lem}\label{lem:d}
For any diagonal matrix $D=\mathrm{diag}\{d_i\}>0$, consider the transformed system of (\ref{eq:Axb}):
\begin{align}
\tilde{A}\tilde{x}=\tilde{b}  \label{eq:Axb1}
\end{align}
with $\tilde{A} = D^{-1}AD$, $\tilde{x}=D^{-1}x$ and $\tilde{b}=D^{-1}b$.  Then, applying Algorithm~\ref{alg1} to (\ref{eq:Axb1}) yields
\begin{align*}
\tilde{a}_{i\rightarrow j}^{(k)} &= a_{i\rightarrow j}^{(k)}; \\
\tilde{b}_{i\rightarrow j}^{(k)} &= d_i^{-1} b_{i\rightarrow j}^{(k)}; \\
\tilde{a}_{i}^{(k)} &= a_{i}^{(k)}; \\
\tilde{b}_{i}^{(k)} &= d_i^{-1} b_{i}^{(k)}.
\end{align*}
In particular, the diagonal transformation does not affect the convergence properties of Algorithm~\ref{alg1}. 
\end{lem}

\begin{IEEEproof}
The verification is direct by noting $\tilde{a}_{ij}= d_i^{-1}a_{ij}d_j$ and $\tilde{b}_i = d_i^{-1}b_i$. 
\end{IEEEproof}

A direct implication of Lemma~\ref{lem:d} is the simple bound below, which resembles the bound in \cite{Roy1} under a somewhat different diagonal dominance condition. 
\begin{cor}\label{cor:1}
Suppose $A$ is generalised diagonally dominant. Define $R=\{r_{ij}\}= I-A_d^{-1}A$ and $\bar{R}=\{|r_{ij}|\}$, where $A_d=\mathrm{diag}\{a_{ii}\}$. Then,
\begin{align}
\frac{|x_i^{(k)}-x_i^{\star}|}{u_i}\le \rho^{k+1} \|x^{\star}\|_u, \label{eq:rateboundsimple}
\end{align} 
where $\rho<1$ is the spectral radius of $\bar{R}$, and $u>0$ is the eigenvector of $\bar{R}$ corresponding to $\rho$, i.e., $\bar{R}u=\rho u$. 
\end{cor} 

\begin{IEEEproof}
The facts that $\rho<1$ and $\bar{R}u=\rho u$ with $u>0$ come from the assumption that $A$ is generalised diagonally dominant; see, e.g., \cite{Mailoutov}. From $\bar{R}u= \rho u$, we get $\sum_{j\in i} a_{ii}^{-1}|a_{ij}|u_j = \rho u_i$ for all $i$. 
 Consider the positive diagonal matrix $U=\mathrm{diag}\{u_i\}$. Then, the above implies $U^{-1}AU$ is diagonally dominant because this is the same as $\rho a_{ii}u_i = \sum_{j\ne i} |a_{ij}|u_j$. Taking $\varrho_i = \rho$, we have $\varrho_i<1$ and $\varrho_i a_{ii}u_i \ge \sum_{j\ne i} |a_{ij}|u_j$ for all $i$, i.e., $A$ is $D$-scaled diagonally dominant with $D=U$. Now apply Theorem~\ref{thm:1}. The recursion of $\eta_{i\rightarrow j}^{(\ell)}$ in (\ref{eq:beta_l}) implies that  
 \begin{align*}
 \eta_{i\rightarrow j}^{(\ell)} &= \varrho_i \frac{\sum_{v\in N_i\backslash j} \Lambda_{v\rightarrow i}^{(\ell-1)} \eta_{v\rightarrow i}^{(\ell-1)}}{|a_{ij}|d_j(1-\varrho_i\varrho_j)+\sum_{v\in N_i\backslash j} \Lambda_{v\rightarrow i}^{(\ell-1)}}\\
 &\le \varrho_i \frac{\sum_{v\in N_i\backslash j} \Lambda_{v\rightarrow i}^{(\ell-1)} \eta_{v\rightarrow i}^{(\ell-1)}}{\sum_{v\in N_i\backslash j} \Lambda_{v\rightarrow i}^{(\ell-1)}}\\
 &\le \rho\max_{v\in N_i\backslash j}  \eta_{v\rightarrow i}^{(\ell-1)}. 
 \end{align*}
Carrying out the above process repeatedly for $\ell-1, \ell-2$, etc. and using $\rho_{i\rightarrow j}^{(0)}=\rho_i=\rho$ for all $i,j$, we will eventually get   
\begin{align*}
 \eta_{i\rightarrow j}^{(\ell)} &\le 
 \rho^{\ell+1}.
 \end{align*} 
 Using~(\ref{eq:rate}) yields 
 \begin{align*}
\frac{|x_i^{(k)}-x_i^{\star}|}{u_i}\le \varrho_i\frac{\sum_{v\in N_i} \Lambda_{v\rightarrow i}^{(k-1)}\eta_{v\rightarrow i}^{(k-1)}}{\sum_{v\in N_i}\Lambda_{v\rightarrow i}^{(k-1)}} \|x^{\star}\|_u\le \rho^{k+1} \|x^{\star}\|_u,
\end{align*}
which is  (\ref{eq:rateboundsimple}). 
\end{IEEEproof}

Define the {\em asymptotic convergence rate} $\lambda_{\infty}$ as 
\begin{align}
\lambda_{\infty} &= \mathrm{arg}\inf_{\lambda>0}\left \{\lambda:  \lim_{k\rightarrow \infty} \frac{\|x^{(k)} - x^{\star}\|}{\lambda^k} \le C \|x^{\star}\|\right \}  \label{eq:asymptotic_rate}
\end{align}
for a constant $C>0$. 
We see from corollary above that $\lambda_{\infty}\le \rho$, i.e., $\rho$ serves as a simple {\em upper bound} for $\lambda_{\infty}$. (Note that the choice of norm does not affect the asymptotic convergence rate.)

It is tempting to conjecture that $\rho$ is the asymptotic convergence rate of Algorithm~\ref{alg1} for a generalised diagonally dominant matrix $A$. But Example 1 in Section~\ref{sec6:1} will show that this is not the case. That is, the convergence rate bound in Theorem~\ref{thm:1} can be tighter than $\rho$ asymptotically by taking the diagonal transformation matrix $D$ other than $U$. 

\begin{figure}[ht]
\begin{picture}(240,135)
\put(54,124){\circle{6}}
\put(57,123){\line(1,-1){39}} 
\put(51,123){\line(-1,-1){39}} 
\put(54,121){\line(0,-1){38}} 
\put(54,80){\circle{6}}
\put(12,81){\circle{6}}
\put(96,81){\circle{6}}
\put(54,77){\line(0,-1){38}}
\put(56,38){\line(1,1){40}} 
\put(52,38){\line(-1,1){40}} 
\put(54,36){\circle{6}}
\put(58,30){$5$}
\put(58,126){$1$}
\put(58,75){$3$}
\put(17,75){$2$}
\put(99,75){$4$}

\put(176,134){\circle{6}}
\put(179,133){\line(1,-1){39}} 
\put(173,133){\line(-1,-1){39}} 
\put(176,131){\line(0,-1){38}} 
\put(176,90){\circle{6}}
\put(134,91){\circle{6}}
\put(218,91){\circle{6}}
\put(176,87){\line(0,-1){25}}
\put(134,88){\line(0,-1){25}}
\put(218,88){\line(0,-1){25}}
\put(134,60){\circle{6}}
\put(176,60){\circle{6}}
\put(218,60){\circle{6}}
\put(180,136){$1$}
\put(180,85){$3$}
\put(139,85){$2$}
\put(221,85){$4$}
\put(139,60){$5$}
\put(180,60){$5'$}
\put(222,60){$5''$}
\put(133,57){\line(-3,-5){12}}
\put(135,57){\line(3,-5){12}}
\put(175,57){\line(-3,-5){12}}
\put(177,57){\line(3,-5){12}}
\put(217,57){\line(-3,-5){12}}
\put(219,57){\line(3,-5){12}}
\put(121,34){\circle{6}}
\put(147,34){\circle{6}}
\put(163,34){\circle{6}}
\put(190,34){\circle{6}}
\put(205,34){\circle{6}}
\put(231,34){\circle{6}}
\put(121,31){\line(0,-1){24}}
\put(147,31){\line(0,-1){24}}
\put(163,31){\line(0,-1){24}}
\put(190,31){\line(0,-1){24}}
\put(205,31){\line(0,-1){24}}
\put(231,31){\line(0,-1){24}}
\put(121,4){\circle{6}}
\put(147,4){\circle{6}}
\put(163,4){\circle{6}}
\put(190,4){\circle{6}}
\put(205,4){\circle{6}}
\put(231,4){\circle{6}}
\put(110,32){$4'$}
\put(137,32){$3'$}
\put(167,32){$2'$}
\put(177,32){$4''$}
\put(209,32){$3''$}
\put(235,32){$2''$}
\put(110,2){$1'$}
\put(136,2){$1''$}
\put(166,2){$1'''$}
\put(179,2){$1`$}
\put(209,2){$1``$}
\put(235,2){$1```$}
\put(141,115){$\varrho_2$}
\put(123,75){$\varrho_5$}
\put(115,47){$\varrho_4$}
\put(143,47){$\varrho_3$}
\put(110,17){$\varrho_1$}
\put(136,17){$\varrho_1$}

\put(165,115){$\varrho_3$}
\put(165,75){$\varrho_5$}
\put(158,47){$\varrho_2$}
\put(185,47){$\varrho_4$}
\put(154,17){$\varrho_1$}
\put(181,17){$\varrho_1$}

\put(202,115){$\varrho_4$}
\put(207,75){$\varrho_5$}
\put(200,47){$\varrho_3$}
\put(227,47){$\varrho_2$}
\put(195,17){$\varrho_1$}
\put(221,17){$\varrho_1$}
\end{picture}
  \caption{The unwrapped tree in Fig.~\ref{fig:2}, with gains $\varrho_i$}\label{fig:2-1}
\end{figure}
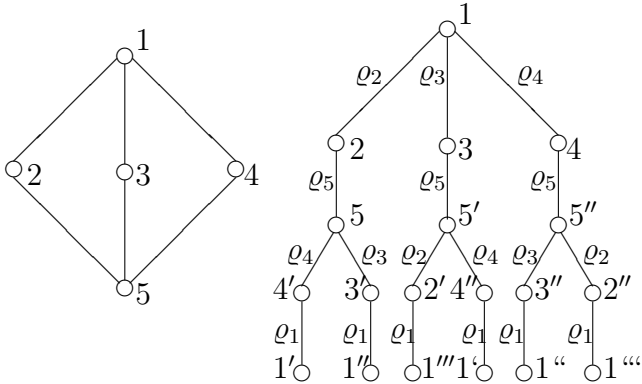 

In the following, we give another simple bound for the asymptotic convergence rate which can be tighter than $\rho$. 

\begin{defn}\label{def:3}
For the induced graph $G=\{V, E\}$, a {\em non-reversal path} $p=\{i_0,i_1, i_2, \ldots, i_{k}\}$ is a path such that $i_{\ell+2}\ne i^{(\ell)}$ for all $\ell=0, 1, \ldots, k-2$.  
For each $i\in V$, denote by $P_i^{(k)}$ the set of all non-reversal paths of length $k$ in $G$ terminated at node $i$. Define, for each path $p=\{i_0, i_1, i_2, \ldots, i_{k-1}, i\}\in P_i^{(k)}$, the {\em path gain} $g(p) = \varrho_{i_0}\varrho_{i_1}\ldots \varrho_{i_{k-1}}$. The path $p$ is called a {\em simple loop} if $i_0=i$ and $i_{\ell}\ne i$ for all $\ell=1, 2,\ldots, k-1$. Further define the {\em loop gain per node} for each simple loop $p$ as 
\begin{align}
\lambda(p)=(g(p))^{1/k} \label{eq:path0}
\end{align}
and denote the {\em maximum loop gain per node}, $\lambda_{\star}$, as the largest loop gain per node among all the simple loops in $G$, i.e., 
\begin{align*}
\lambda_{\star} &= \max \{\lambda(p): \ p \mathrm{\ is\ a\ simple\ loop\ in\ } G\}.
\end{align*}
\end{defn}

We see (\ref{eq:beta_l}) that $\eta_{i\rightarrow j}^{(\ell)}$ is formed by a weighted sum of all $\eta_{v\rightarrow i}^{(\ell-1)}$, $v\in N_i\backslash j$, then multiplied by $\varrho_i$. This is more clearly depicted in Fig.~\ref{fig:2-1}, a reproduced version of the unwrapped tree in Fig.~\ref{fig:2} with {\em gains} $\varrho_i$ attached on each edge. This implies that the bound (\ref{eq:rate}) can be interpreted as a weighted sum of the path gains for all the (non-reversal) paths from the leaf layer to the root node. For the example in Fig.~\ref{fig:2-1}, we see 6 paths, which are $(1',4',5,2,1)$, $(1'', 3', 5, 2,1)$, etc., and their corresponding path gains are $\varrho_1\varrho_4\varrho_5\varrho_2$, $\varrho_1\varrho_3\varrho_5\varrho_2$, etc. Stated more formally, (\ref{eq:rate}) can be re-expressed as 
\begin{align}
\frac{|x_i^{(k)} - x_i^{\star}|}{d_i} &\le \varrho_i\|x^{\star}\|_d\sum_{p\in P_i^{(k)}} w(p)g(p),\label{eq:rate2}
\end{align}
where $w(p)>0$ is the weight of $p$ with $\sum_{p\in P_i^{(k)}} w(p)\le1$. 

Based on the above observations, we are ready to give the next simple bound for $\lambda_{\infty}$. 

\begin{thm}\label{thm:2}
Suppose $A\in \mathbb{R}^{n\times n}$ is weakly $D$-scaled diagonally dominant for some diagonal matrix $D>0$. Then, 
\begin{align}
\lambda_{\infty}\le \lambda_{\star}<1. 
\label{eq:rate7}
\end{align}
\end{thm}

\begin{IEEEproof}
Take any terminating (root) node $i\in V$ and consider an arbitrary non-reversal path $p=\{i_0, i_1, \ldots, i_{k-1}, i\}$ in $G$ with length $k> n$ (where $n$ is the number of nodes in $G$). It is obvious that $p$ contains at least one loop because $k>n$. Split $p$ into three parts: $p_1=\{i_0, i_1, \ldots, i_{k_1}\}$, $p_2=\{i_{k_1+1}, \ldots, i_{k_2}\}$ and $p_3=\{i_{k_2+1}, \ldots, i_{k-1}, i\}$ in their connecting order, such that $p_2$ is a loop and $p_1$ and $p_3$ together do not have repeating nodes. It is clear that $p_1$ and $p_3$ together have less than $n$ nodes. Denote the length of $p_2$ by $\tilde{k}$. It is obvious that if we replace every node $j$'s $\varrho_j$ in $p_2$ with $\lambda_{\star}$, the loop gain per node will not decrease and it will become $\lambda_{\star}$.  Therefore, 
\begin{align*}
(g(p))^{1/k} &= [(g(p_1)g(p_3))^{1/k}(g(p_2))^{\frac{1}{k}-\frac{1}{\tilde{k}}}](g(p_2))^{1/\tilde{k}} \\
&=[(g(p_1)g(p_3))^{1/k}(g(p_2))^{\frac{\tilde{k}-k}{k\tilde{k}}}](g(p_2))^{1/\tilde{k}} \\
&\le [(g(p_1)g(p_3))^{1/k}(g(p_2))^{\frac{\tilde{k}-k}{k\tilde{k}}}]\lambda_{\star}.
\end{align*}
As $k\rightarrow \infty$, the term in $[\ \ ]$ approaches 1. Hence, 
\begin{align*}
\lim_{k\rightarrow \infty}  (g(p))^{1/k} \le \lambda_{\star}.
\end{align*}
Applying the above to (\ref{eq:rate2}), we get
\begin{align*}
\lim_{k\rightarrow \infty}\frac{|x_i^{(k)}-x_i^{\star}|}{d_i\lambda_{\star}^k} &\le \lim_{k\rightarrow \infty}\varrho_i \|x^{\star}\|_d\sum_{p\in P_i^{(k)}} w(p) \frac{g(p)}{\lambda_{\star}^k} \\
&\le \lim_{k\rightarrow \infty} \varrho_i \|x^{\star}\|_d \sum_{p\in P_i^{(k)}} w(p) =\varrho_i\|x^{\star}\|_d.
\end{align*}
Hence, 
\begin{align*}
\lim_{k\rightarrow \infty} \frac{\|x^{(k)}-x^{\star}\|_d}{\lambda_{\star}^k} \le (\max_{i}\varrho_i) \|x^{\star}\|_d.
\end{align*} 
Since the asymptotic convergence rate is independent of choice of the norm $\|\cdot\|$, the above means that $\lambda_{\star}\ge \lambda_{\infty}$. 

Finally, we check $\lambda_{\star}<1$. Consider any simple loop $p$ in $G$. If $p$ has an even number of nodes, then every pair of adjacent nodes $i$ and $j$ in $p$ have $\varrho_i\varrho_j<1$ by assumption, so $g(p)<1$. If $p$ has an odd number of nodes, we traverse the loop by starting with the node $i_0$ with maximum $\varrho_{i_0}$. Again, since every pair of adjacent nodes $i$ and $j$ in $p$ have $\varrho_i\varrho_j<1$, we have $g(p)< \varrho_{i_k}$, where $i_k$ is the last node in $p$ before returning back to $i_0$, for which we must have $\varrho_{i_k}<1$ because it is adjacent to $i_0$ with the maximum $\varrho_{i_0}$. Therefore, $g(p)<1$ for every simple loop $p$, which means that $\lambda_{\star}<1$. 
\end{IEEEproof}

The only remaining question is how tight the bound $\lambda_{\star}$ is in comparison with $\rho$ in Corollary~\ref{cor:1}. For this question, we note that $\lambda_{\star}$ depends on the given diagonal matrix $D$ and we have the following simple answer. 

\begin{cor}\label{cor:2}
Suppose $A$ is generalised diagonally dominant. Then, 
\begin{align}
\lambda_{\infty}\le \min_{D} \lambda_{\star} \le \rho, \label{eq:rate10}
\end{align} 
where the diagonal $D>0$ is such that $A$ is weakly $D$-scaled diagonally dominant.  
\end{cor}
\begin{IEEEproof}
The result is easily established by noting that, if we take $D=U=\mathrm{diag}\{u_i\}$ with $u>0$ being the eigenvector corresponding to $\rho$ for $\bar{R}$ as in Corollary~\ref{cor:1}, we have $\varrho_i=\rho<1$ for all $i$, leading to $\lambda_{\star}=\rho$.  
\end{IEEEproof}

\begin{rem}\label{rem:3}
Note that the proof for $\lambda_{\star}$ being an upper bound of the convergence rate for Algorithm~\ref{alg1} is derived from the general bound in Theorem~\ref{thm:1}. This means that the general bound in Theorem~\ref{thm:1} is tighter than the bound $\min_D \lambda_{\star}$, which in turn is tighter than $\rho$. Since $\rho$ is the known bound for Gaussian BP \cite{Roy1}, the bounds in Theorems~\ref{thm:1}-\ref{thm:2} and Corollaries~\ref{cor:1}-\ref{cor:2} are tighter than the known bound in the literature, even in the case of Gaussian BP.  
\end{rem}

\begin{rem}
We conclude this section by explaining briefly how our bounds on the convergence rate can be applied. 
Firstly, these bounds are useful in determining the performance of the algorithm when compared with other distributed and iterative algorithms. An example will be given for comparison with the classical Jacobi method (see Example 3 in the next section). Secondly, these bounds can be used to determine {\em a priori} how many iterations are required to reach a given level of accuracy. Taking the simple bound (\ref{eq:rateboundsimple}) for example, if $\|x^{(k)}-x^{\star}\|_u/\|x^{\star}\|_u \le \epsilon$ is required for a given $\epsilon>0$, the required number of iterations $k$ can be easily bounded by solving $\rho^{k+1} = \epsilon$. Other bounds can be used in a similar way. 
\end{rem} 

\section{Illustrating Examples}\label{sec6}

This section illustrates our results using several examples. 

\subsection{Example 1: Single-Loop System}\label{sec6:1}

\begin{figure}[ht]
\begin{center}
\begin{picture}(210,135)
\put(54,124){\circle{6}}
\put(57,123){\line(1,-2){42}} 
\put(51,123){\line(-1,-2){42}} 
\put(8,36){\circle{6}}
\put(100,36){\circle{6}}
\put(11,36){\line(1,0){85}} 
\put(58,126){$1$}
\put(07,25){$2$}
\put(98,25){$3$}

\put(164,124){\circle{6}}
\put(167,123){\line(1,-2){42}} 
\put(161,123){\line(-1,-2){42}} 
\put(164,121){\line(0,-1){38}} 
\put(164,80){\circle{6}}
\put(118,36){\circle{6}}
\put(210,36){\circle{6}}
\put(164,77){\line(0,-1){38}}
\put(167,36){\line(1,0){40}} 
\put(161,36){\line(-1,0){40}} 
\put(164,36){\circle{6}}
\put(164,25){$3$}
\put(168,126){$1$}
\put(168,75){$2$}
\put(117,25){$4$}
\put(208,25){$5$}
\end{picture}
  \caption{Single-Loop Graph in Example 1; Double-Loop Graph in Example 2}\label{fig:s20}
  \end{center}
\end{figure}
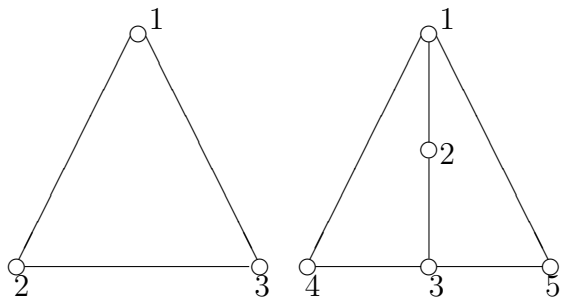

\begin{figure}
\begin{center}
  \includegraphics[width=8.5cm]{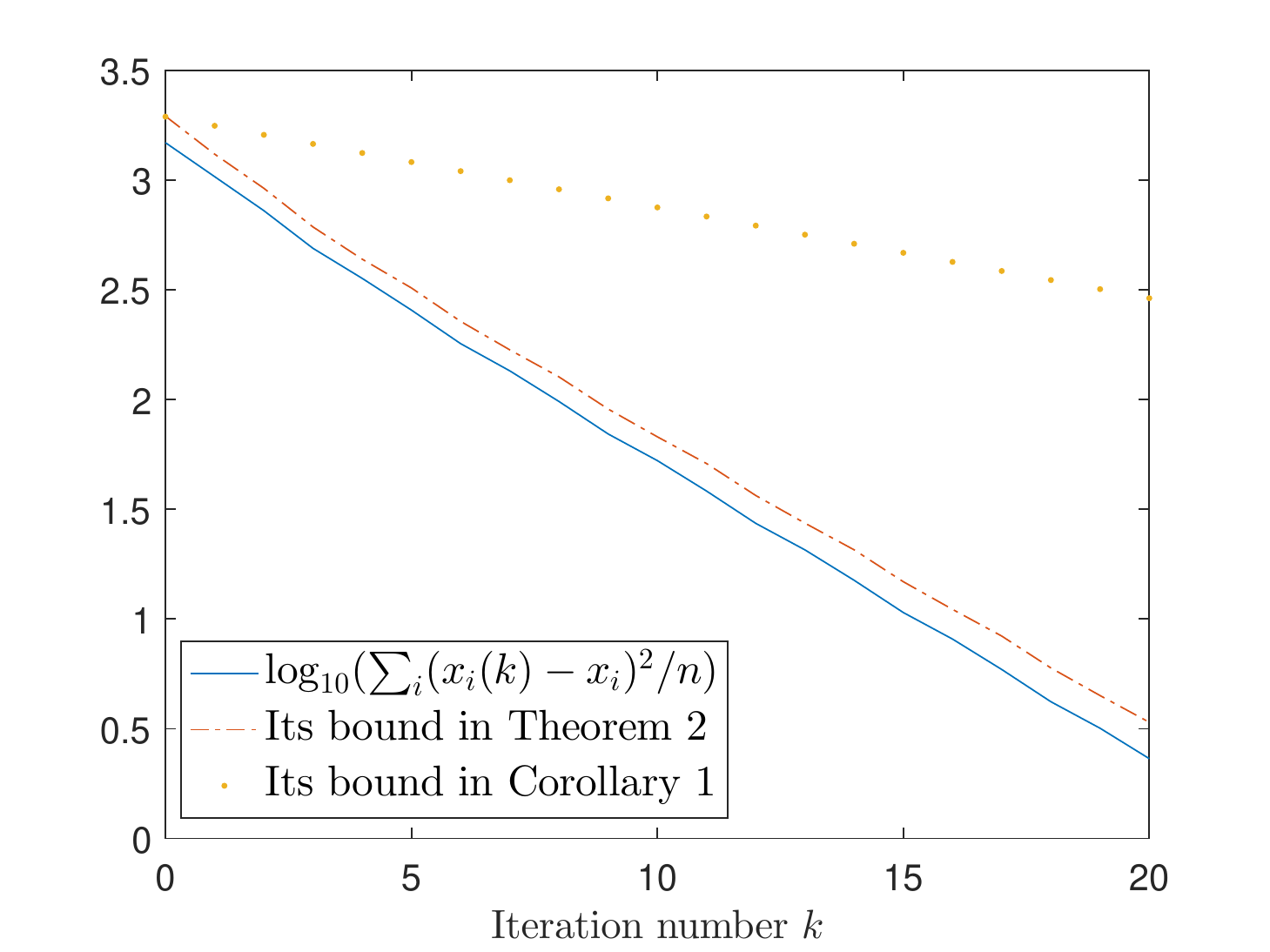}  
\end{center}
  \caption{Convergence of the 3-node Single-Cycle Graph for Example 1}\label{fig:s1}
\end{figure}

Consider a single-loop system in Fig.~\ref{fig:s20} with 
\begin{align*}
A&=\left [\begin{array}{ccc}1 &-0.72 &-0.6\\-0.1&1&-0.375\\-0.7&-0.5&1\end{array} \right ] 
\end{align*}
and $b_i=i$ for all $i$. It is easy to see that $A$ is not diagonally dominant. By taking $D=\mathrm{diag}\{1,0.565,0.98\}$, it can be verified that $D^{-1}AD$ is diagonally dominant. Hence, $A$ is $D$-scaled diagonally dominant (and generalised diagonally dominant). The single loop is simply $\{1,2,3,1\}$. For this $D$, we can compute that $\varrho_1=0.9978, \varrho_2=0.8308, \varrho_3=0.9975$. It is computed using Theorem~\ref{thm:2} that $\lambda_{\star}=(\varrho_1\varrho_2\varrho_3)^{1/3}=0.95386$. Also computed directly from $A$ is $\rho =0.9535$.  Instead of showing how $x_i^{(k)}-x_i^{\star}$ decreases over $k$ for each $i$, we plot in Fig.~\ref{fig:s1} a congregated curve of
\begin{align}
\log_{10}\frac{1}{n}\|x^{(k)}-x^{\star}\|^2= \log_{10}\frac{1}{n}\sum_{i} (x_i^{(k)}-x_i^{\star})^2 \label{eq:slope}
\end{align}
along with its bound given by Theorem~\ref{thm:1} and that by $\rho$ (i.e., Corollary~\ref{cor:1}).
It is estimated from Fig.~\ref{fig:s1} that the slope of (\ref{eq:slope}) is approximately $-0.1650$, which corresponds to a convergence rate of $10^{-0.1650/2}= 0.8270$. This confirms that both Theorem~\ref{thm:2} and Corollary~\ref{cor:1} are correct for this example.

\subsection{Example 2: Double-Loop System}

\begin{figure}
\begin{center}
  \includegraphics[width=8.5cm]{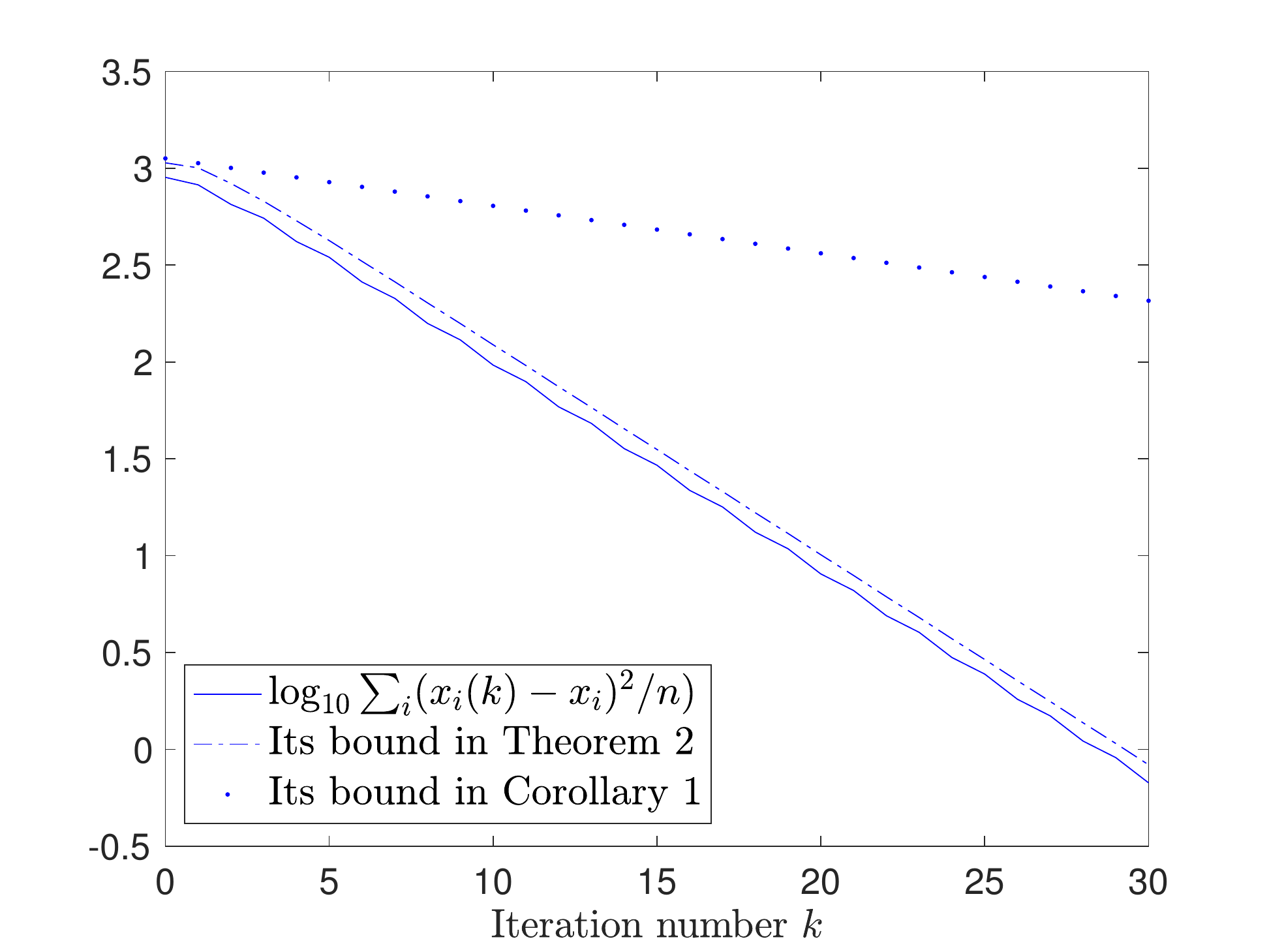}
\end{center}
  \caption{Convergence of the Double-Cycle Graph for Example 2}\label{fig:s2}
\end{figure}

Consider a double-loop system in Fig.~\ref{fig:s20} with 
\begin{align*}
A&=\left [\begin{array}{ccccc}1 & 0.29 & 0 & 0.32 & 0.35 \\
                                        0.51& 1&0.48&0&0\\
                                        0&0.3&1&0.32&0.35\\
                                        0.52&0&0.46&1&0\\
                                        0.44&0&0.53&0&1\end{array} \right ]
\end{align*}
and $b_i=i$ for all $i$. Also take $D=I$.  There are three simple loops $p_1=\{1,2,3,4,1\}$, $p_2=\{1,2,3,5,1\}$ and $p_3=\{1,4,3,5,1\}$. We have $\varrho_1=0.96, \varrho_2=0.99, \varrho_3=0.97, \varrho_4=0.98, \varrho_5=0.97$. It is computed that $\lambda_{\star}=0.9749$ and $\rho =0.9722$.  The simulation results are shown in Fig.~\ref{fig:s2} and the slope of (\ref{eq:slope}) is approximately $-0.1033$, corresponding to a convergence rate of $10^{-0.1033/2}= 0.8879$. 

\subsection{Example 3: 13-node Loopy Graph}
\begin{figure}
\begin{center}
  \includegraphics[width=7cm]{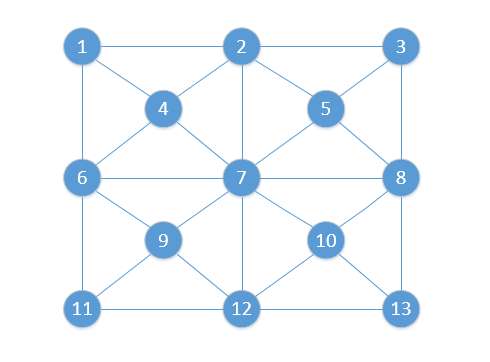}
\end{center}
  \caption{A 13-node Loopy Graph for Example 3}\label{fig:S.5}
\end{figure}

This example considers a 13-node loopy graph shown in Fig.~\ref{fig:S.5}. The matrix $A$ has $a_{ii}=|N_i|$ and the non-zero $a_{ij}$ randomly chosen from $(-1.2, -0.2)$, and $b_i=i$.  Also take $D=I$. For a particular realisation of $A$, it is verified that $A$ is diagonally dominant and its $\rho=0.9586$.  The logarithmic error $\log_{10}(\|\hat{x}^{(k)} - x\|^2/n)$ is plotted in Fig.~\ref{fig:S.6}, along with its bound given by Theorem~\ref{thm:1} and that by $\rho$ (i.e., Corollary~\ref{cor:1}). 

\begin{figure}
\begin{center}
  \includegraphics[width=9.5cm]{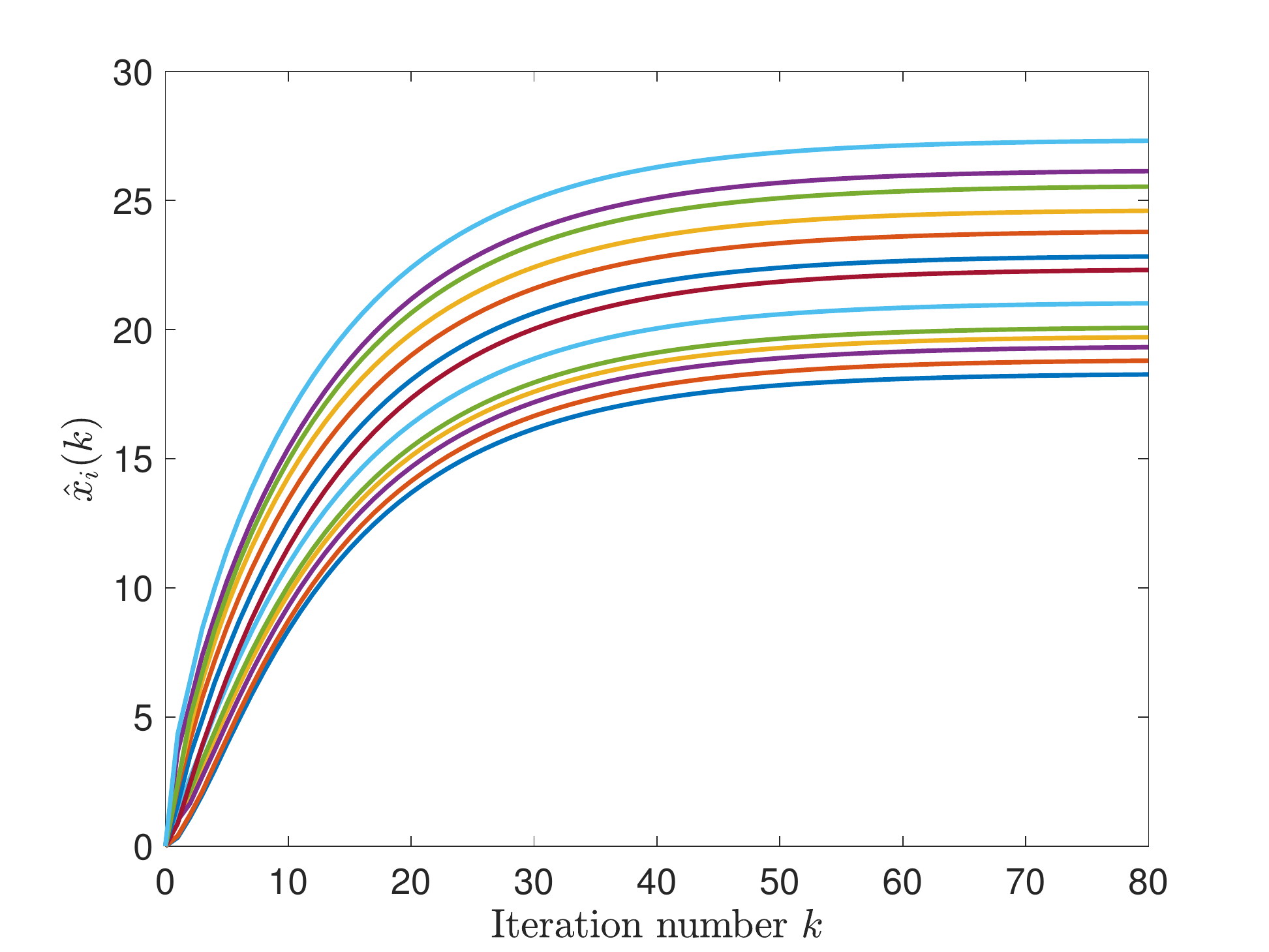}
\end{center}
  \caption{Convergence of the Jacobi method on loopy graph}\label{fig:S.7}
\end{figure}

In addition, we compare Algorithm~\ref{alg1} with the classical Jacobi method \cite{Horn,Saad} which is a common iterative algorithm for solving linear systems. The simulated result for Algorithm~\ref{alg1} is shown in Fig.~\ref{fig:S.6}. For 100 iterations, the error is converged down to approximately $0.8\times10^{-4}$. Fig.~\ref{fig:S.7} shows the simulated result for the Jacobi method, which has a considerably slower convergence rate, with an error of approximately 0.02 after 100 iterations. It is known \cite{Horn,Saad} that the Jacobi method has a convergence rate equal to the spectral radius of $R$. Thus, this simulation shows that Algorithm~\ref{alg1} has a faster convergence rate than that of the Jacobi method.

\begin{figure}
\begin{center}
  \includegraphics[width=8.5cm]{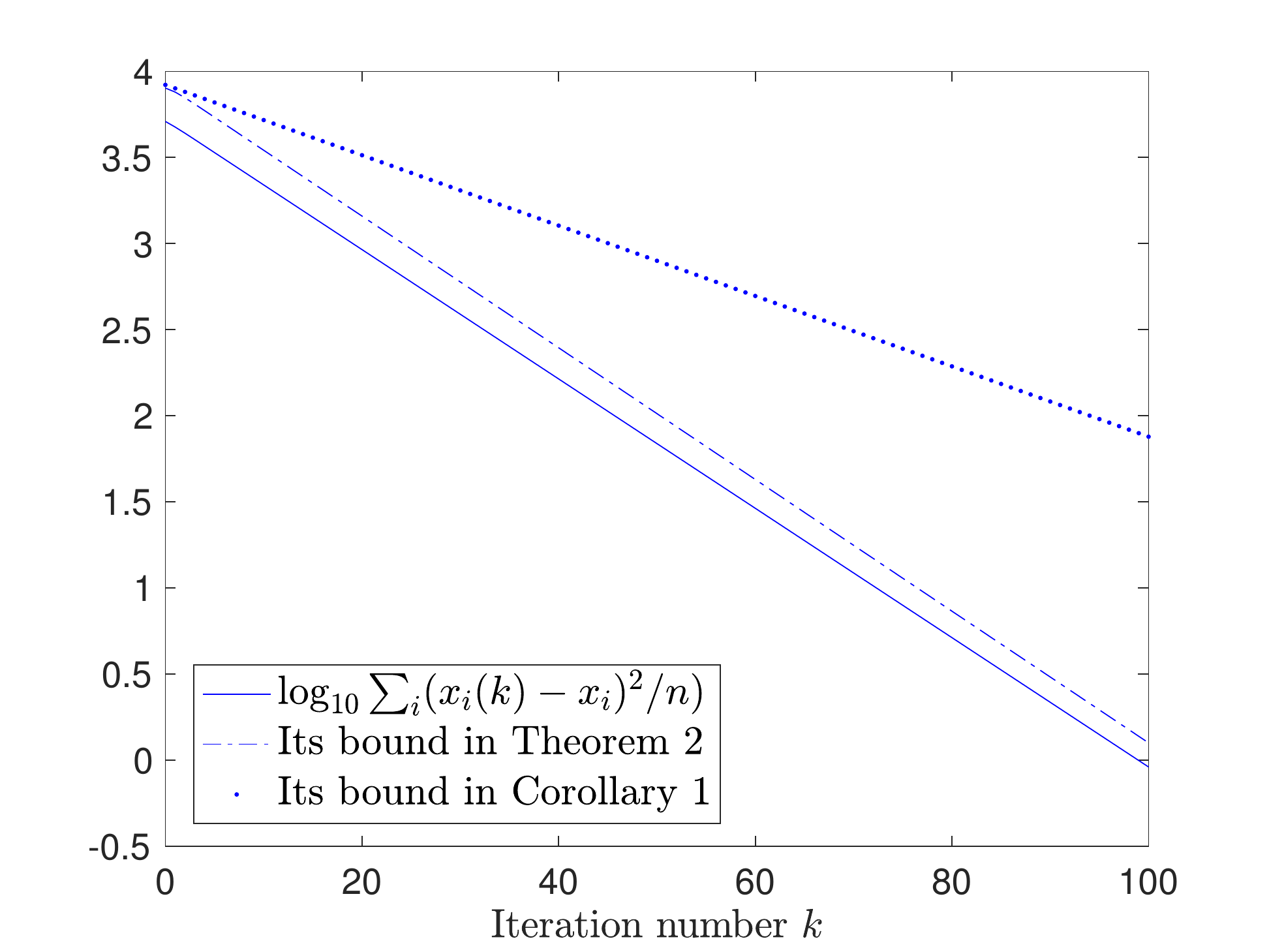}
\end{center}
  \caption{Convergence of the 13-node System in Example 3}\label{fig:S.6}
\end{figure}

\begin{figure}
\begin{center}
   \includegraphics[width=9.5cm]{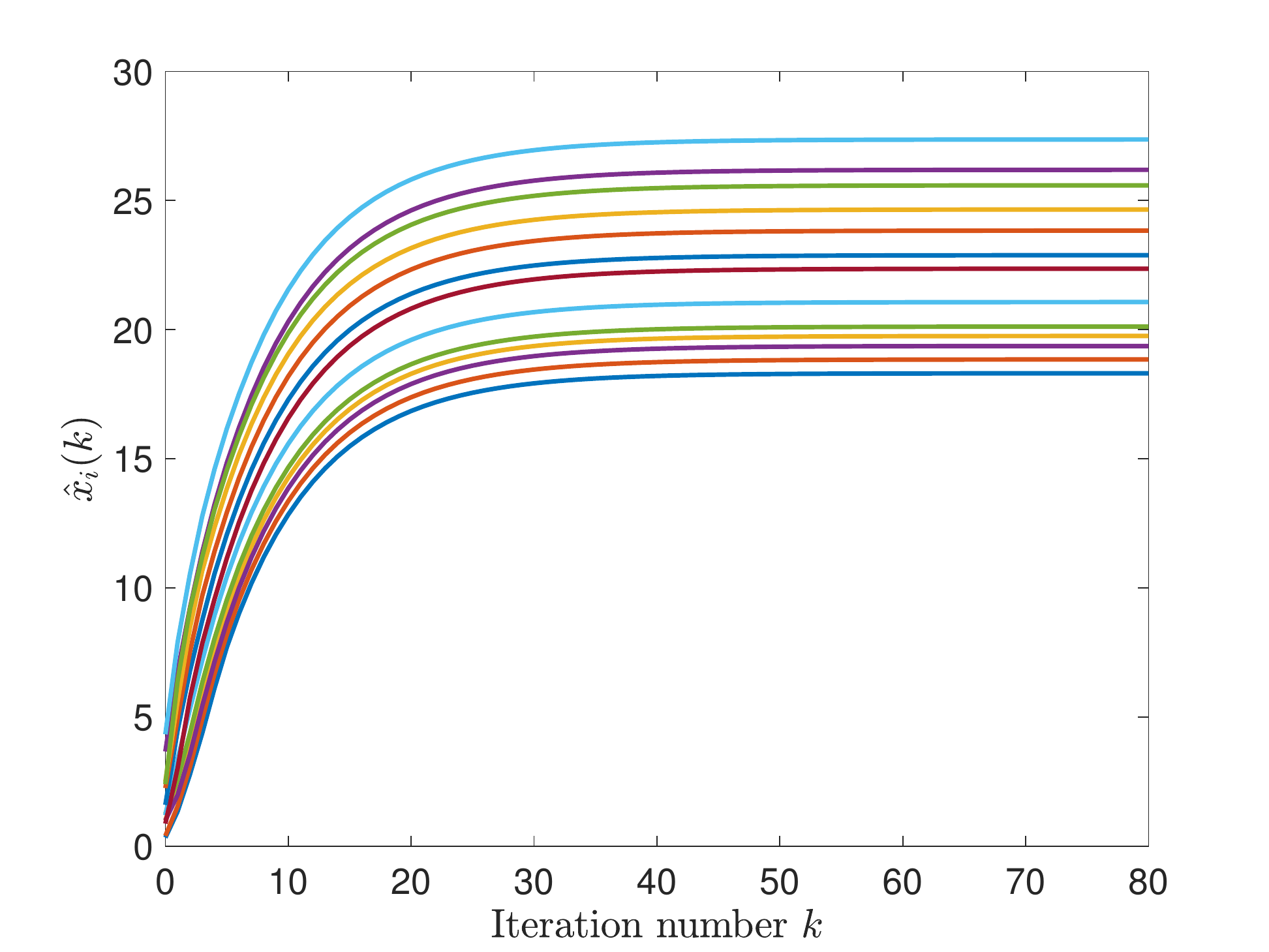}
\end{center}
  \caption{Convergence of Algorithm~\ref{alg1} on loopy graph}\label{fig:S.6}
\end{figure}

\subsection{Example 4: Large-Scale System}
\begin{figure}
\begin{center}
  \includegraphics[width=8.5cm]{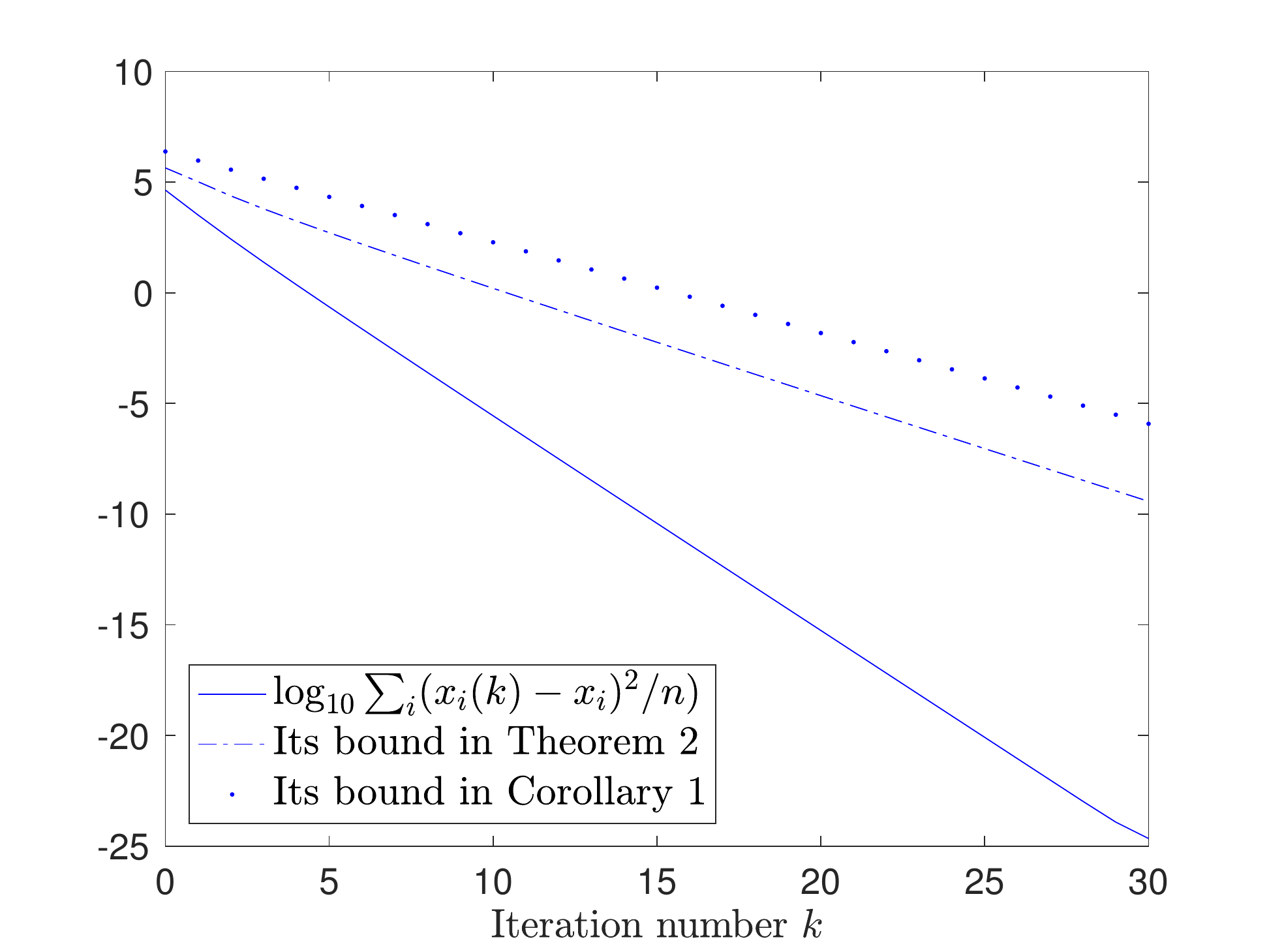}
\end{center}
  \caption{Convergence of the 1000-node System in Example 4}\label{fig:S.11}
\end{figure}

This example involves a randomly connected 1000-node  loopy graph shown in Fig.~\ref{fig:S.5}. The circles indicate the nodes and the curves indicated the edges. The matrix $A$ has $a_{ii}=1$ for all $i$ with the average number of edges for each $i$ to be about  7.772. The non-zero off-diagonal terms $a_{ij}$ to be random with 80\% probability to be positive and each row's absolute sum is 0.4 on average. Again, $b_i=i$ for all $i$.  It is computed that $\rho=0.9356$. The logarithmic error $\log_{10}(\|\hat{x}^{(k)} - x\|^2/n)$ is plotted in Fig.~\ref{fig:S.11} along with  its bound given by Theorem~\ref{thm:1} and that by $\rho$ (i.e., Corollary~\ref{cor:1}). Since many off-diagonal terms of $A$ are positive, the bound given by Theorem~\ref{thm:1} is not as tight as for the previous two examples.

\begin{figure}
\begin{center}
  \includegraphics[width=9cm]{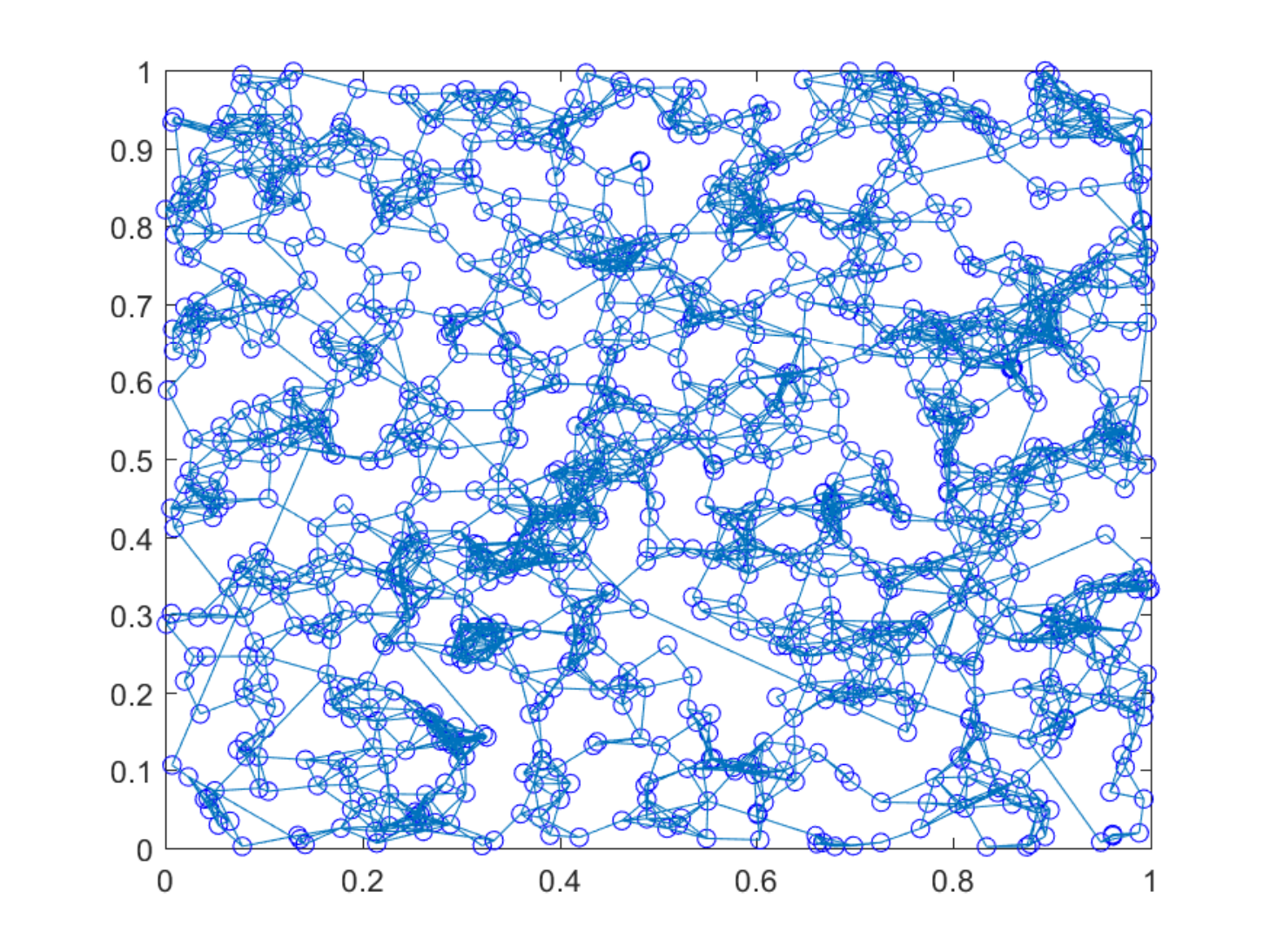}
  \end{center}
  \caption{A 1000-node Random Loopy Graph for Example 4}\label{fig:S.10}
\end{figure}

\section{Conclusions}\label{sec7}

In this paper, we have studied the convergence rate of a message-passing distributed algorithm for solving linear systems.  Under the assumption of generalised diagonal dominance, the convergence rate of the distributed algorithm is shown to be explicitly related to the diagonal dominance properties of the system (Theorems~\ref{thm:1}-\ref{thm:2} and Corollaries~\ref{cor:1}-\ref{cor:2}).  Due to the fact this algorithm is more general than the Gaussian BP algorithm in the sense that it deals with both symmetric and non-symmetric matrices, the results in this paper also apply to the Gaussian BP algorithm.  

The distributed algorithm studied in this paper belongs to the so-called {\em synchronous algorithms}, meaning that every node needs to update their variables (or messages) simultaneously in each iteration. The Gaussian BP algorithm can also be implemented in an asynchronous matter (called asynchronous Gaussian BP), and it is known that convergence properties exist under the generalised diagonal dominance assumption \cite{Su,Su1}. Likewise, the message-passing algorithm, Algorithm~\ref{alg1}, can be implemented in an asynchronous matter. It is expected that our convergence rate analysis approach can be applied in the asynchronous case as well, which is one of the future tasks. 

\section*{Appendix}

\begin{lem}\label{lem:d0}
Suppose $A\in\mathbb{R}^{n\times n}$ is weakly $D$-scaled diagonally dominant for a given diagonal matrix $D>0$. Then, $A$ is generalised diagonally dominant. 
\end{lem}

\begin{IEEEproof}
We only consider the case with at least one $\varrho_i\ge1$ because otherwise $A$ is obviously generalised diagonally dominant. Define $U=\{i:\varrho_i\ge1\}$. For each $i\in U$, define $\bar{\rho}_i = \varepsilon+ \max_{j\in N_i}\varrho_j$ for some sufficiently small $\varepsilon>0$ such that $\bar{\rho}_i<1$ and $\tilde{\rho}_i=\varrho_i\bar{\rho}_i<1$. This can always be done because $\varrho_i\varrho_j<1$ for every $j\in N_i$. For each $i\not\in U$, define $V_i=\{j: j\in N_i, j\in U\}$ and $\tilde{\rho}_i=\varrho_i\breve{\rho}_i^{-1}$, where
\begin{align*}
\breve{\rho}_i &= \left \{\begin{array}{lll} 1 & \mathrm{if} & V_i=\phi (\mathrm{empty}) \\
\min_{j\in V_i}\bar{\rho}_j & \mathrm{if} & V_i\ne \phi. \end{array}\right.
\end{align*} 
It is clear that $\breve{\rho}_i\le 1$. 

Next, define a new diagonalising matrix $\tilde{D}=\mathrm{diag}\{\tilde{d}_i\}$ with
\begin{align*}
\tilde{d}_i &= \left \{ \begin{array}{lll} \bar{\rho}_i^{-1}d_i & \mathrm{if} & i\in U\\ d_i & \mathrm{if} & i\not\in U. \end{array}\right.
\end{align*}
We claim that $\tilde{D}^{-1}A\tilde{D}$ is diagonally dominant. Indeed, consider
\begin{align*}
\Delta_i &= \tilde{\rho}_ia_{ii}\tilde{d}_i - \sum_{j\in N_i} |a_{ij}|\tilde{d}_j. 
\end{align*}
For each $i\in U$, we know that any $j\in N_i$ must be such that $j\not\in U$ (due to $\varrho_i\varrho_j<1$), hence, $\tilde{d}_j=d_j$ and 
\begin{align*}
\Delta_i &= \bar{\rho}_i\varrho_i a_{ii}d_i\bar{\rho}_i^{-1} - \sum_{j\in N_i} |a_{ij}|d_j = 0.
\end{align*}
For each $i\not\in U$, we have
\begin{align*}
\Delta_i &= \varrho_i\breve{\rho}_i^{-1}a_{ii}d_i - \sum_{j\in N_i}|a_{ij}|\tilde{d}_j.
\end{align*}
We consider two cases of $j\in N_i$. Case 1: $j\not\in U$, for which $\tilde{d}_j=d_j$. Case 2: $j\in U$, for which $\tilde{d}_j = d_j\bar{\rho}_j^{-1}$. It follows that 
\begin{align*}
\Delta_i &= \breve{\rho}_i^{-1}\hspace{-1mm}\left ( \varrho_ia_{ii}d_i - \hspace{-4mm}\sum_{j\in N_i, j\not\in U} |a_{ij}|d_j \breve{\rho}_i - \hspace{-4mm}\sum_{j\in N_i, j\in U} |a_{ij}|d_j \breve{\rho}_i\bar{\rho}_j^{-1}\hspace{-1mm}\right ).
\end{align*}
Recall $\breve{\rho}_i\le 1$. Also, under $i\not\in U, j\in N_i, j\in U$, we have $j\in V_i$ by the definition of $V_i$, i.e., $V_i\ne \phi$. Hence, $\breve{\rho}_i \le \bar{\rho}_j$ from the definition of $\breve{\rho}_i$. Applying these facts to $\Delta_i$ above, we get, for any $i\not\in U$, 
\begin{align*}
\Delta_i &\ge  \breve{\rho}_i^{-1}\left ( \varrho_ia_{ii}d_i - \hspace{-4mm}\sum_{j\in N_i, j\not\in U} |a_{ij}|d_j -\hspace{-4mm} \sum_{j\in N_i, j\in U} |a_{ij}|d_j\right )=0.
\end{align*}
Therefore, $\Delta_i\ge0$ for all $i$.  It remains to confirm that $\tilde{\rho}_i<1$ for all $i$. This is obvious for the case of $i\in U$ by the choice of $\varepsilon$ earlier. For the case of $i\not\in U$, $\tilde{\rho}_i=\varrho_i\breve{\rho}_i^{-1}$. If $V_i=\phi$, we have $\breve{\rho}_i=1$ and thus $\tilde{\rho}_i=\varrho_i<1$. If $V_i\ne \phi$, 
\begin{align*}
\tilde{\rho}_i &= \frac{\varrho_i}{\min_{j\in V_i} \bar{\rho}_j} 
=\max_{j\in V_i} \frac{\varrho_i}{\bar{\rho}_j} = \max_{j\in N_i, j\in U} \frac{\varrho_i}{\bar{\rho}_j}.
\end{align*}
Note that for each $j\in N_i$ with $j\in U$, $\bar{\rho}_j>\max_{v\in N_j}\varrho_v \ge \varrho_i$. So, $\tilde{\rho}_i<1$ in this case as well.
We have confirmed that $\tilde{\rho}_i<1$ for all $i$. Therefore, $\tilde{D}^{-1}A\tilde{D}$ is diagonally dominant. 
\end{IEEEproof}

\end{document}